\documentclass[11pt,preprintnumbers,amsmath,amssymb,nofootinbib,groupedaddress,superscriptaddress]{revtex4}

\usepackage{graphicx}
\usepackage{dcolumn}
\usepackage{epsfig}
\usepackage[dvips]{color}
\usepackage{bm}
\usepackage{mathrsfs}
\usepackage{booktabs}
\usepackage{amsmath}
\usepackage{amssymb}

\allowdisplaybreaks[4]

\begin{document}

\preprint{MPP-2011-40}
\preprint{TUM-HEP-806/11}

\title{Light Neutralino in the MSSM: a playground for dark matter, flavor physics and collider experiments}

\author{Lorenzo Calibbi}\email{calibbi@mppmu.mpg.de}
\affiliation{Max-Planck-Institute f\"ur Physik (Werner-Heisenberg-Institute), D-80805 M\"unchen, Germany}

\author{Toshihiko Ota}\email{toshi@mppmu.mpg.de}
\affiliation{Max-Planck-Institute f\"ur Physik (Werner-Heisenberg-Institute), D-80805 M\"unchen, Germany}

\author{Yasutaka Takanishi}\email{yasutaka.takanishi@ph.tum.de}
\affiliation{Physik-Department, Technische Universit\"at M\"unchen, D-85748 Garching, Germany}

\date{\today}

\begin{abstract}
We investigate the constraints to the light neutralino dark matter
scenario in the minimal supersymmetric standard model 
from available experimental observations such as
decays of $B$ and $K$ meson, relic dark matter abundance,
and the search for neutralino and Higgs production at colliders.
We find that two regions of the MSSM parameter space fulfill 
all the constraints: a fine-tuned strip with large $\tan\beta$ 
where the lightest neutralino can be as light as 8 GeV, and a low
$\tan\beta$ region providing a neutralino mass larger than
16 GeV. 
The large $\tan\beta$ strip, which can be compatible with recently
reported signals from direct detection experiments,
can be fully tested by means of low-energy observables and,
in particular, by $B_s\to \mu\mu$ and Higgs bosons searches at the LHC
within the upcoming months.

\end{abstract}

\maketitle


\section{Introduction}

The existence of dark matter (DM) has been established by
plenty of cosmological observations, and its total abundance in the universe 
has been evaluated in the last decade with high precision by the 
WMAP experiment \cite{Dunkley:2008ie,Komatsu:2010fb}.
A possible interpretation of such observations relies on the 
existence of a new stable particle species, the so-called WIMP 
(weakly interacting massive particle). However, the nature and the properties
of this new particle are far from being established.
Numerous candidates have been proposed from the theoretical side,
and numerous attempts have been made to detect it both directly 
and indirectly. 

Some of the direct detection experiments (in which nucleon recoil by
invisible incoming particles is measured) have recently reported possible signals of
DM, while others have found no excess above the background.  The situation has not been
settled yet (for some recent discussions, see e.g.
Refs.~\cite{Kopp:2009qt,Regis:2010ay,Schwetz:2010gv}).  
Besides the long-standing DAMA~\cite{Bernabei:2008yi,Bernabei:2010mq} evidence for an annual
modulated signal, claims for signals have been recently made by 
CoGeNT~\cite{Aalseth:2010vx} and CRESST~\cite{NewCRESST:2011}.  What is intriguing is that all
these experimental results could be explained by a rather light WIMP
with a sizeable spin-independent scattering cross-section with
nucleons~\cite{Hooper:2010uy}:
\begin{equation}
       m_{\chi} \sim 7\div10 \text{ GeV}\,,\quad \sigma_{\chi N}^{\text{SI}} \sim 10^{-41}\div 10^{-40} \text{ cm$^{2}$}\,. 
\label{Eq:cogent}
\end{equation}

On the other hand, 
the most extensively studied framework for physics beyond the standard model (SM)
is represented by supersymmetry (SUSY) and, in particular, by the
minimal supersymmetric extension of the SM (so-called MSSM), 
with the lightest neutralino as a candidate of dark matter, in case R-parity is conserved. 
It is therefore natural and compelling to ask whether the MSSM neutralino
can account for the DM properties suggested by the direct detection 
experiments, as given in Eq.~(\ref{Eq:cogent}).
This question has been recently addressed by several 
collaborations~\cite{Kuflik:2010ah,Feldman:2010ke,Vasquez:2010ru,Fornengo:2010mk} 
(for previous works on light neutralinos, see \cite{Hooper:2002nq,Dreiner:2009ic}).
However, these works do not agree in the conclusions.
In particular, the authors of Ref.~\cite{Vasquez:2010ru} performed a fit
requiring the WMAP constraint on the DM relic density together with several 
indirect constraints on the model parameters to be satisfied. 
They concluded that the lightest neutralino mass must be larger than 28 GeV, thus
excluding that the properties summarized in Eq.~(\ref{Eq:cogent}) 
can be accounted for by the MSSM lightest neutralino.
On the contrary, the authors of Ref.~\cite{Fornengo:2010mk}
claimed the viability of the parameter regime with a light neutralino 
($\gtrsim 7$ GeV) and the direct detection cross-section as large as $\sim \mathcal{O}(10^{-41})$ cm$^{2}$.

In this paper, we study the MSSM parameter space to re-consider 
the compatibility of the lightest neutralino with the properties suggested by the 
direct detection experiments and shed light on the conflict mentioned above. 
We restrict ourselves on the particle content of 
the MSSM, i.e. we do not consider possible variations or extensions of the minimal model 
(such as the NMSSM), which have been recently studied in this context by different collaborations~\cite{Das:2010ww,Gunion:2010dy,Draper:2010ew,Vasquez:2010ru,Kappl:2010qx}.
The key points of our analysis are summarized in the following.
\begin{itemize}
\item We study in detail the low-energy constraints, which challenge a light neutralino scenario, with a 
particular emphasis on the processes that have no strong dependence on the SUSY parameters but only on the Higgs sector
parameters. The most relevant observables turn out to be the $B\to\tau\nu$ decay and the Kaon physics observable 
$R_{\ell 23}$, extracted from the decay $K\to\mu\nu$, which has not considered in previous works on light neutralino dark matter.
\item We perform a numerical scan of the SUSY parameter space, without adopting any high-energy relation among the parameters,
which are instead treated as low-energy free parameters, and we then identify the regions fulfilling all the relevant constraints.
\item We study the impact of LEP and Tevatron Higgs searches on the light neutralino parameter space.
\item We analyze the specific consequences for the Higgs sector, the low-energy processes, the SUSY parameter space and direct DM
searches, discussing the prospects for experimentally probing the light neutralino scenario.
\end{itemize}
Besides a quite different treatment of the experimental contraints, with the inclusion of more observables in the analysis 
(such as $R_{\ell 23}$), the phenomenological analysis of the 
paramater space compatible with light neutralino DM represents the main new feature of the present work with respect
to the previous related literature \cite{Kuflik:2010ah,Feldman:2010ke,Vasquez:2010ru,Fornengo:2010mk}. In particular, 
as we will see, this study will allow us to identify several independent ways to test light neutralino DM scenarios in the near
future.

The rest of the paper is organized as follows. The main constraints are discussed in section \ref{sec:lowE};
in section \ref{sec:numerical} the numerical analysis of the parameter space is presented and the 
phenomenological consequences are discussed in section \ref{sec:prospects}. In section \ref{sec:direct}, 
the impact of direct DM search experiments is discussed. Finally, our findings are summarized in section \ref{sec:summary}. 

\section{Constraints to light neutralino dark matter scenarios}
\label{sec:lowE}

Within the MSSM, a neutralino as light as in Eq.~(\ref{Eq:cogent}) requires
$M_1 \ll M_2,\,\mu$, where $M_1$ and $M_2$ are respectively the $U(1)$ and $SU(2)$ gaugino mass parameters
and $\mu$ is the Higgs bilinear mixing parameter of the superpotential. In fact, $M_2$ and $\mu$ enter the chargino
mass matrix and are therefore bounded
to be $\gtrsim 90$ GeV, by the LEP limit on chargino masses. As a result, the lightest neutralino is mostly Bino. 
As usual in the MSSM, the Bino-like DM (especially if light, as in the case we are considering)
is thermally overproduced in the early universe. Therefore, to obtain the appropriate relic abundance,
an efficient annihilation process is necessary.
Since an efficient slepton mediated annihilation would require sleptons with masses smaller than the LEP limit~\cite{Dreiner:2009ic},
for a neutralino mass in the range we are interested in, the only way to enhance the neutralino annihilation cross-section 
is through the mediation of a CP-odd Higgs boson~\cite{Bottino:2002ry}.
This requires the CP-odd Higgs to be as light as $m_{A} \sim 100$ GeV and a somewhat large value of $\tan\beta$~\cite{Bottino:2002ry,Bottino:2004qi,Bottino:2008mf,Fornengo:2010mk}.
The entire spectrum of the extended Higgs sector of the MSSM is therefore required to lie around 100 GeV 
(we recall for instance the tree level relation $m^2_{H^\pm}=m^2_A + m^2_W$). 
This choice of the parameters causes several phenomenological difficulties. Indeed, CP-odd and charged Higgs exchanges
contribute to several rare decays. Therefore, a light Higgs sector can induce large deviations from the SM,
especially in the large $\tan\beta$ regime \cite{Buras:2002vd}. For recent discussions,  see e.g.~\cite{Isidori:2006pk,Barenboim:2007sk,Eriksson:2008cx,Altmannshofer:2009ne,Altmannshofer:2010zt}.
We can conveniently group the most relevant observables in two categories.
\begin{itemize}
 \item[(i)] The processes, which mainly depend on the Higgs sector parameters (i.e. $m_A$ and $\tan\beta$) and have a dependence on the 
   other SUSY parameters only through threshold corrections to the Yukawa couplings. This group includes the following decays:
     $B\to\tau\nu$, $B\to D \tau\nu$, $D_s \to \ell\nu$, $K\to \mu\nu$.
  \item[(ii)] The processes which get Higgs-mediated contributions but have in addition a non-trivial dependence 
           on the SUSY spectrum and the SUSY parameters. For our discussion, the most relevant observables of this kind 
           are $B_s\to\mu\mu$ and $b\to s \gamma$.
\end{itemize}
In the case of the processes of this second group, an appropriate choice of the SUSY parameters can avoid unacceptably large
deviations from the SM expectations. As it is well known, this is the case of $b\to s \gamma$, whose charged Higgs contribution
can be compensated by a sizeable stop-chargino contribution.   
On the other hand, the processes of group (i) provide strong constraints to the possible values of $m_A$ and $\tan\beta$, which
challenge a light neutralino scenario almost independently on the details of the SUSY spectrum.
Therefore, let us start reviewing the processes of group (i).

\subsection{Group (i) constraints}

The charged Higgs boson ($H^\pm$) mediates the $B \rightarrow \tau \nu$ decay at tree-level.
Remarkably, the charged Higgs contribution has opposite sign with respect 
to the SM contribution. The resulting deviation from the SM prediction
can be then expressed as following \cite{Hou:1992sy,Akeroyd:2003zr,Isidori:2006pk}:
\begin{align}
R_{B\tau\nu}
\equiv
\frac{{\rm BR} (B \rightarrow \tau \nu)}
{{\rm BR} (B \rightarrow \tau \nu)_{\rm SM}}
\simeq
\left[
1 -
\frac{m_{B}^{2}}{m_{H^{\pm}}^{2}}
\frac{\tan^{2} \beta}{1 + \epsilon \tan \beta}
\right]^{2}.
\label{eq:btn}
\end{align}
where $m_B$ is the mass of the $B^\pm$ meson ($\simeq 5.3$ GeV) and 
$\epsilon$ accounts for the non-holomorphic SUSY 
threshold corrections to the down quark Yukawas, 
while we neglect the corresponding leptonic correction
for the moment.
 
Due to the dependence of $R_{B\tau\nu}$ on $\tan\beta$, 
the experimentally allowed range of $R_{B\tau\nu}$
will clearly identify two possible ranges of $\tan \beta$ for 
a given value of $m_{H^\pm}$: either a small/moderate $\tan \beta$ regime,
where BR($B \rightarrow \tau \nu$) is decreased with respect to the
SM prediction only to a certain extent, or a large $\tan \beta$ regime,
for which the charged Higgs contribution is approximately twice the SM one,
so that the difference between them still provides a value
for $R_{B\tau\nu}$ within the experimental range.
This second possibility is however strongly challenged by other 
processes, for which the Higgs-mediated contribution increases
with $\tan\beta$.

The most relevant example is $K\to \ell\nu$. In order to
reduce theoretical uncertainties, it is convenient to consider
the following quantity~\cite{Antonelli:2008jg}:
\begin{equation}
R_{\ell 23} \equiv \left|\frac{V_{us}(K\to \ell\nu)}{V_{us}(K\to \pi\ell\nu)} \times
\frac{V_{ud}(\beta{\rm~decay})}{V_{ud}(\pi\to\ell\nu)}\right|\,,
\end{equation}
where $V_{us}$ and $V_{ud}$ represent the values of the CKM matrix entries
as extracted from the processes indicated in the parentheses.
This quantity depends mainly on the mass of the charged Higgs 
boson and $\tan \beta$ as well. It is therefore complementary to
$B \rightarrow \tau \nu$ in constraining the parameter space,
especially for the larger $\tan \beta$ 
region among two allowed regions by $B \rightarrow \tau \nu$.  
The analytic formula is given by~\cite{Antonelli:2008jg}:
\begin{align}
R_{\ell 23}
\simeq&
\left|
1 
-
\frac{m_{K}^{2}}{m_{H^{\pm}}^{2}}
\left[
1 - 
\frac{m_{d}}{m_{s}}
\right]
\frac{\tan^{2} \beta}{1 + \epsilon \tan \beta}
\right|,
\label{eq:rl23}
\end{align}
where $m_{K} = 0.494$ GeV and the quark mass ratio 
$m_{d}/m_{s}$ takes a value between 1/22 and 1/17~\cite{Nakamura:2010zzi}.

For a large value of $\tan\beta$, the 3-body decay $B \rightarrow D \tau \nu$ 
may deviate from the SM prediction as well. 
The complete formula for $B \rightarrow D \tau \nu$
is more involved than the 2-body decays which we have 
considered so far. A compact approximated 
expression, given in Ref.~\cite{Kamenik:2008tj}, reads
\begin{align}
R_{D\ell\nu}
\equiv
\frac{{\rm BR}(B \rightarrow D \tau \nu)}{{\rm BR}(B \rightarrow D e
 \nu)}
=
(0.28 \pm 0.03)
\times
\left[
1 + 1.38(6) 
C_{\rm NP}^{\tau} 
+
0.88(4)
(C_{\rm NP}^{\tau})^{2}
\right],
\label{eq:xdln}
\end{align}
where
\begin{align}
C_{\rm NP}^{\tau}
\equiv
-
\frac{m_{b} m_{\tau}}{m_{H^{\pm}}^{2}}
\frac{\tan^{2} \beta}{1 + \epsilon \tan \beta}.
\end{align}

The authors of Ref.~\cite{Akeroyd:2009tn} 
pointed out the possible relevance of the process $D_{s} \rightarrow \tau \nu$.
The analytic formula given in Ref.~\cite{Akeroyd:2009tn} reads
\begin{align}
{\rm BR}(D_{s} \rightarrow \tau \nu)
=\!
\frac{G_{F}^{2}}{8\pi}
|V_{cs}|^{2} f_{D_{s}}^{2}
m_{\tau}^{2} 
m_{D_{s}} \tau_{D_{s}}
\left[
1 - \frac{m_{\tau}^{2}}{m_{D_{s}}^{2}}
\right]^{2}
\!\left[
1 + \frac{1}{m_{c} + m_{s}}
\frac{m_{D_{s}}^{2}}{m_{H^{\pm}}^{2}}
\left(
m_{c} - \frac{m_{s} \tan^{2} \beta}{1 + \epsilon  \tan \beta}
\right)
\right]^{2}\!,
\label{eq:dtn}
\end{align}
where $m_{D_s}$, $\tau_{D_s}$, $f_{D_s}$ are respectively the mass,
the life-time and the decay constant of the $D^\pm_s$ meson.

We can now use the expressions given above, in order to 
constrain the parameters $\tan\beta$ and $m_A$ (or equivalently $m_{H^\pm}$). 
From the experimental data, it is possible to extract the following 95\% C.L. ranges for 
the considered observables:
\begin{center}
\begin{tabular}{cc}
$0.52 < R_{B\tau\nu} < 2.61 $ & \cite{Altmannshofer:2009ne,Altmannshofer:2010zt} \\
$0.985 < R_{\ell23}(K\to\mu\nu)  < 1.013 $ & \cite{Antonelli:2010yf} \\
$0.151 < R_{D\ell\nu} < 0.681 $ & \cite{Aubert:2007dsa} \\
$4.7\times 10^{-2} < {\rm BR}(D_{s} \rightarrow \tau \nu) < 6.1\times 10^{-2} $ & \cite{Onyisi:2009th,Alexander:2009ux}
\end{tabular} 
\end{center}
\begin{figure}[t]
\centering
\includegraphics[width=0.50\textwidth]{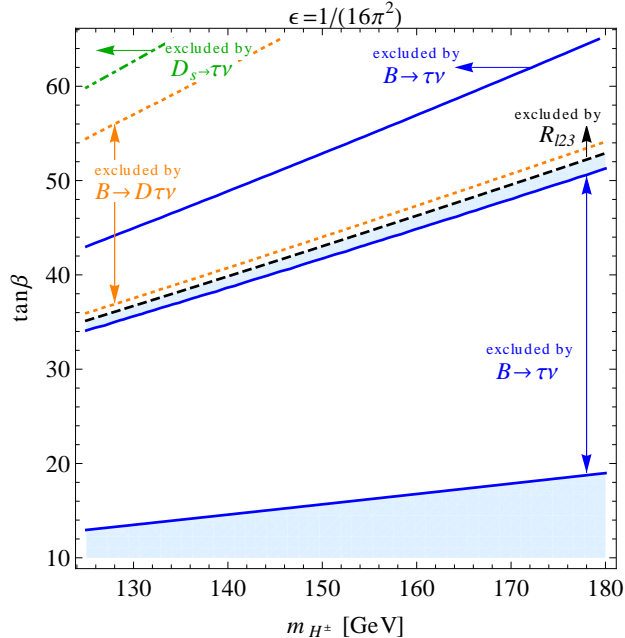}
\caption{Constraint to the mass of the charged Higgs boson and $\tan \beta$ 
from the group (i) observables. 
The lines indicate the boundaries of the exclusion area (95 \% C.L.)
by $B \rightarrow \tau \nu$ (blue), $R_{\ell 23}$ (black dashed), 
$B \rightarrow D \tau \nu$ (orange),
and $D_{s} \rightarrow \tau \nu$ (green).
The common allowed region by the four observables is shown with 
the light blue shade.
Here the SUSY threshold correction $\epsilon$ is taken at the reference 
value $1/(16 \pi^{2}).$\label{fig:const}
}
\end{figure}

The constraints on the $m_{H^\pm}$-$\tan\beta$ plane resulting from the expressions in 
Eqs.~(\ref{eq:btn}, \ref{eq:rl23}, \ref{eq:xdln}, \ref{eq:dtn}) together with 
the experimental ranges reported above are shown in Fig.~\ref{fig:const}. The light blue
shaded regions survive all the constraints listed above.
As already mentioned, $B\to\tau\nu$ excludes a wide portion of the plane, leaving 
unconstrained the low $\tan\beta$ region and a band with $\tan\beta \gtrsim 30$ 
(namely, the region between the two blue lines in the figure).
This band corresponds to the case of a large charged Higgs contribution 
(larger about 1.5 times than the SM one), such that
the SM contribution is overcompensated and $R_{B\tau\nu}$ ``re-enters'' the 
experimental range $0.52 < R_{B\tau\nu} < 2.61 $. This is the reason why a large value 
of $\tan\beta$ is required, and the band is almost excluded by 
other observables, whose deviation from the SM also increases with $\tan\beta$, in particular
$B \rightarrow D \tau \nu$ (orange line) and $R_{\ell 23}$ (black dashed line). 
From the figure, we see that the most constraining observable is $R_{\ell 23}$.
Indeed, only a quite thin strip in the plane remains viable in the large
$\tan\beta$ regime. Besides that, all the constraints can be evaded only
for small values of $\tan\beta$ ($\lesssim 20$ in the region displayed in the figure).

Let us now comment about the mild dependence on the SUSY spectrum of these results.
The plot has been made for the illustrative value $\epsilon = 1/(16\pi^2)$, while in the
next section the SUSY thresholds will be computed numerically (including the leptonic ones).
However, it is remarkable that, as long as the same $\epsilon$ enters
Eqs.~(\ref{eq:btn}, \ref{eq:rl23}, \ref{eq:xdln}, \ref{eq:dtn}), the bounds
on the $m_A$-$\tan\beta$ plane shift in the same way. The thin surviving strip,
for instance, moves upwards by increasing $\epsilon$, but it does not get shrunk 
(nor disappear). The situation could change if we consider different values for $\epsilon$
entering the expressions for $R_{B\tau\nu}$, $R_{\ell 23}$, as indeed it is the case
when third generation squarks are not degenerate with the first generations. Still,
the strip cannot be excluded, unless third generation squarks are consistently 
lighter than the first generation ones.

Finally, let us notice that previous works~\cite{Eriksson:2008cx,Mahmoudi:2010xp} claimed that $R_{\ell 23}$ 
completely excludes the large $\tan\beta$ region left unconstrained by $B \rightarrow \tau \nu$, 
since they considered as allowed range $0.990 < R_{\ell23}< 1.018$~\cite{Antonelli:2008jg},
which have been recently updated in Ref.~\cite{Antonelli:2010yf}. We also find that using such
previous bound no viable large $\tan\beta$ region would remain. 

\subsection{Group (ii) constraints}
Here we qualitatively review the requirements to the SUSY parameters 
from the bounds of the group (ii) observables.
Later, we will treat them quantitatively in our numerical study.
Differing from the group (i) processes which are discussed in the previous 
subsection, the observables categorized into group (ii) depend strongly on 
the SUSY spectrum and parameters.

In MSSM, there are three major contributions to $b\rightarrow s\gamma$ 
process~\cite{Bertolini:1990if}, (a) the SM contribution from a $W$-top loop, (b) the two Higgs doublet model (THDM) 
contribution from a charged Higgs-top loop, whose sign is the same as (a), 
and (c) the SUSY stop-chargino contribution.
In the non-SUSY THDM in which there are only two contributions (a) and
(b), $b\rightarrow s\gamma$ alone constrains strongly the mass of the charged 
Higgs boson ($\gtrsim 300$ GeV~\cite{Ciuchini:1997xe}), almost independently of $\tan\beta$.
However, this constraint to $m_{H^{\pm}}$ is considerably weakened in the
MSSM by a possible cancellation between the contribution (a)+(b) and the SUSY 
contribution (c)~\cite{Barbieri:1993av,Okada:1993sx,Garisto:1993jc}. As discussed above, we have to set the mass of the 
CP-odd Higgs to be light 
to reproduce the correct amount of relic density. This parameter choice enhances the
contribution (b). In order to diminish $b\rightarrow s\gamma$ and allow
the light CP-odd and charged Higgs bosons regime, 
the above mentioned compensation must be invoked.
A sufficiently large chargino contribution requires relatively 
light stop and chargino.
Since the contribution is proportional to the soft SUSY breaking 
trilinear coupling $A_{t}$ and $\tan \beta$,
these parameters should not take too small values. Moreover, 
$A_t$ and $\mu$ must have opposite sign, so that the chargino and
charged Higgs contributions are opposite in sign too.
However, these requests raise another well-constrained 
observable, BR($B_{s} \rightarrow \mu\mu$). This is because
the dominant contribution to this decay is mediate by a neutral Higgs 
($H^0$ or $A$) exchange with an 
effective $b$-$\bar{s}$-Higgs vertex given by a
higgsino-stop loop similar to the one contributing to
$b\rightarrow s\gamma$. 
Therefore, it is necessary to have a balance of these two observables, especially
for large values of $\tan\beta$. In order to do so,
the SUSY parameters have to be carefully chosen and will consequently 
exhibit non-trivial correlations.
For example, since $B_{s} \rightarrow \mu\mu$ is strongly enhanced 
by the sixth power of $\tan\beta$, the value of $A_t$ should be
reduced in the large $\tan\beta$ regime.
A more detailed discussion of the parameter correlation will be presented
in the section for the numerical analysis.

Let us finally recall that in the large $\tan\beta$ regime, 
the non-holomorphic SUSY threshold corrections become important~\cite{Ciuchini:1998xy,Carena:2000uj,Degrassi:2000qf}, 
and we take it into account in our numerical study.

\subsection{Relic density and scattering with nuclei}
\label{Sec:relic}

Before we move on to our numerical results, let us briefly discuss  
the parameter choice to obtain the correct relic density.
Due to the chargino
mass limit from the LEP experiments ($\gtrsim 90$ GeV) the SUSY
parameters $\mu$ (the higgsino mass) and $M_2$ (the Wino mass), the lowest of which
basically sets the chargino mass, 
must be larger than about 100 GeV. Clearly, the condition $M_1\ll M_2, \mu$ should
be satisfied to get a very light lightest neutralino.
From this consideration follows that  $m_{\widetilde{\chi}^{0}_{1}}\simeq M_1$
and the lightest neutralino is mostly Bino.

As previously mentioned, such light neutralino is thermally 
overproduced in the early universe, and an efficient annihilation process 
is necessary to reduce surplus neutralinos and reproduce the DM abundance evaluated
from cosmological observations. 
As discussed in Ref.~\cite{Bottino:2002ry}, within the parameter regime described above,
the main contribution to the neutralino annihilation is due to the $s$-channel 
exchange of a neutral CP-odd Higgs, $A$, and the main channel is consequently 
${\chi}^{0}_{1}{\chi}^{0}_{1}\to b \bar{b}$.
Therefore, in this regime, the relic abundance is essentially controlled 
by the parameters in the Higgs sector.
It is obvious that a light $A$ enhances the annihilation 
process and makes the neutralino relic density smaller.
The process is also affected by the value of $\tan\beta$. This is due to the coupling
of the CP-odd Higgs boson with the $d$-quark pair, $G_{Add} = ({m_d}/{v}) \tan\beta$.
First, since the coupling is proportional to mass of the fermion, 
a bottom quark pair is dominantly created as a result of the annihilation process, as already 
mentioned. Then, we also see that the coupling between 
the CP-odd Higgs boson and bottom quark is proportional to $\tan\beta$.
Consequently, a large value of $\tan \beta$ enhances the annihilation rate
diminishing the neutralino relic density.

A more subtle point is the dependence on the value of the higgsino 
mass $\mu$.
Although the lightest neutralino is Bino-like in our scenario,
the pair-annihilation in Higgs boson requires 
a sufficiently large higgsino components, because 
the interaction between neutralino and CP-odd Higgs boson, 
${\chi}^{0}_{1}{\chi}^{0}_{1}A$,  
originates from the gauge interaction of Higgs fields,
that is, the interaction among Bino, Higgs boson, and higgsino.
It follows from the structure of the neutralino mass matrix
that a smaller value of $\mu$ increases higgsino component of 
the lightest neutralino and amplifies the annihilation process. 
Therefore, in order to satisfy the WMAP bound, a $\mu$ parameter
of the order of 100 GeV will be required.

Of course, the Bino-mass $M_{1}$ is also an important parameter, which 
dominantly determines the mass of the lightest neutralino.
A larger value of $M_{1}$ reduces the initial amount of 
neutralino in thermal production. Hence, 
a heavier Bino does not demand too efficient annihilation
to reproduce the correct relic density (so that $m_A$ and $\mu$
can acquire larger values or $\tan\beta$ can be smaller).

The DM direct detection process,  
the elastic neutralino-nucleon scattering,
is also mediated by Higgs bosons (but the CP-even ones)
in the light Higgs sector regime.
Thus, it has a similar parameter dependence
to the annihilation cross-section, 
with large $\tan \beta$ and small $\mu$ raising up 
the cross-section.
The process is driven by Yukawa interactions,
therefore, heavier quark components in a target nucleon 
are important although their distribution is small. 
As a consequence, $s$-quark matrix element induces a rather large  
uncertainty in the evaluation of the cross-section \cite{Bottino:2008mf,Belanger:2010cd}.
%
\begin{table}[t]
\begin{tabular}{cccc}
\hline \hline
&Observable 
& Allowed range
& References 
\\
\hline
WMAP
&$\Omega_{\rm DM} h^{2}$
 & [0.101,~0.123]
 & 	     \cite{Komatsu:2010fb}
\\
\hline
LEP &
$m_{h}$
 & $> 92.8$ GeV
 & \cite{Nakamura:2010zzi}\footnote{%
	     In the decoupling regime $\sin^{2} (\alpha - \beta)
	     \rightarrow 1$, 
	     the limit to the SM Higgs $M_{H} >114.4$ GeV
	     should be recovered.
	     We will discuss the dependence
	     of the bound on the coupling in section~\ref{sec:prospects}.
}
\\
& $m_{A}$
 & $> 93.4$ GeV
 & \cite{Nakamura:2010zzi}
\\
&$M_{\tilde{\chi}^{+}_{1}}$
 & $> 94$ GeV
 & \cite{Nakamura:2010zzi} 
\\
&$\Gamma(Z\to \tilde{\chi}^{0}_{1}\tilde{\chi}^{0}_{1})$
 & $<3$ MeV
 & \cite{ALEPH:2005ema}  
\\
&$\sigma(e^{+}e^{-} \rightarrow \tilde{\chi}^{0}_{1}
\tilde{\chi}^{0}_{2,3})$ 
 & $<0.1$ pb
 & \cite{Abbiendi:2003sc}
\\
\hline
Group (i)
&$R_{B \tau \nu}$
 &[0.52,~2.61]
 & \cite{Altmannshofer:2009ne,Altmannshofer:2010zt}
\\
&$R_{\ell23}$
 & [0.985,~1.013]
 & \cite{Antonelli:2010yf}
\\
&$R_{D \ell \nu}$
 & [0.151,~0.681]
 & \cite{Aubert:2007dsa}
\\
&${\rm BR}(D_{s} \rightarrow \tau \nu)$
 & [0.047,~0.061]
 & \cite{Onyisi:2009th,Alexander:2009ux,Akeroyd:2009tn}
\\
\hline
Group (ii)
&$\text{BR}(b \rightarrow s \gamma)$
 & $[2.89,~4.21]\times 10^{-4}$
 & 	 \cite{Barberio:2008fa} 
\\
&$\text{BR}(B_{s} \rightarrow \mu^{+} \mu^{-})$
 & $<5.1 \times 10^{-8}$
 & \cite{Abazov:2010fs}
\\
\hline \hline
\end{tabular}
\caption{Summary of the constraints.}
\label{Tab:condition}
\end{table}
%

\section{Numerical analysis of the parameter space}
\label{sec:numerical}

In this section, we present the result of a numerical analysis of the light neutralino parameter space.
Instead of assuming high-energy relations among the parameters (such as gaugino mass unification), we
varied randomly the following set of parameters, defined at low energy:
\begin{equation}
\tan\beta,\quad M_1,\quad M_2,\quad M_3,\quad a_0,\quad \mu,\quad m_A,\quad m_{\tilde q},\quad m_{\tilde \ell}, 
\label{eq:parameters}
\end{equation}
where $m_{\tilde q}$ and $m_{\tilde \ell}$ are common soft SUSY breaking masses for the three generations squarks
and sleptons respectively, $m_A$ is the CP-odd Higgs mass, $M_{i}$ ($i=1,2,3$) are the three gauginos masses and $\mu$ represents
the bilinear Higgs coupling in the superpotential. As in Ref.~\cite{Bottino:2002ry}, $a_0$ parameterizes the trilinear terms, in the
following way:
\begin{equation}
 A_u = a_0 Y_u m_{\tilde q},\quad A_d = a_0 Y_d m_{\tilde q},\quad A_\ell = a_0 Y_\ell m_{\tilde \ell},
\end{equation}
where $Y_u$, $Y_d$ and $Y_\ell$ are the up-quark, down-quark and lepton Yukawa matrices, respectively.
The parameters of Eq.~(\ref{eq:parameters}) have been varied in the following ranges:
\begin{center}
\begin{tabular}{ccc}
 ~$M_1 \in [7,30]$ GeV,~ & ~$M_2 \in [100,600]$ GeV, ~& ~ $M_3 \in [400,1200]$ GeV, ~\\
$m_A \in [90,120]$ GeV, & $\mu \in [100,200]$ GeV, &  $a_0 \in [-2,2]$,  \\
~$m_{\tilde q} \in [400,1200]$ GeV, ~& ~$m_{\tilde \ell} \in [100,1200]$ GeV, ~& ~ $\tan\beta \in [5,50]$.  ~
\end{tabular}
\end{center}
The spectrum, the neutralino relic density $\Omega_{\rm DM} h^{2}$ and the scattering cross-section with nucleons
were computed by means of the {\tt SuSpect} \cite{Djouadi:2002ze} and {\tt micrOMEGAs} codes~\cite{micro}, 
while the low-energy observables using {\tt SuperIso}~\cite{Mahmoudi:2008tp}. 
Unless otherwise specified in the text, we apply to each point of the scan the set of constraints displayed
in Tab.~\ref{Tab:condition}.
\begin{figure}[t]
\centering
\includegraphics[height=0.55\textwidth,angle=-90]{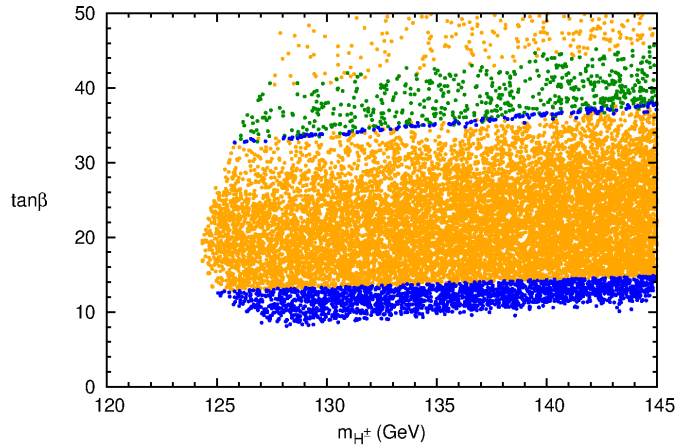}
\caption{$m_{H^\pm}$-$\tan\beta$ plane: the orange (light gray) points satisfy the constraints of Tab.~\ref{Tab:condition}
except the ones of group (i), the green (medium gray) points satisfy in addition the $B\to\tau\nu$ constraint, the blue (dark gray) points
satisfy all the constraints. \label{fig:mHc-tb}}
\end{figure}

Following the discussion of the previous section, let us start looking at the $m_{H^\pm}$-$\tan\beta$ plane. 
The result is showed in Fig.~\ref{fig:mHc-tb}. We first notice that the plot is bounded from below, such
that $\tan\beta\gtrsim 7$. This is a consequence of the upper bound on the neutralino relic density, 
$\Omega_{\rm DM} h^2 < 0.123$. In fact, for smaller values of $\tan\beta$ the neutralino annihilation
cross-section results too low to efficiently decrease the relic abundance. The left part of the plot is
excluded by the LEP limit on $m_A$ ($m_A > 93.4$ GeV). We also notice that the density of points decreases for large values
of $\tan\beta$. This reflects the fact that in such large $\tan\beta$ regime, the other parameters (in particular
$a_0$ and $m_{\tilde q}$) have to be tuned to limited ranges, in order the bounds from $b\to s\gamma$ and
$B_s\to\mu\mu$ to be fulfilled. 
\begin{figure}[t!]
\centering
\includegraphics[height=0.55\textwidth,angle=-90]{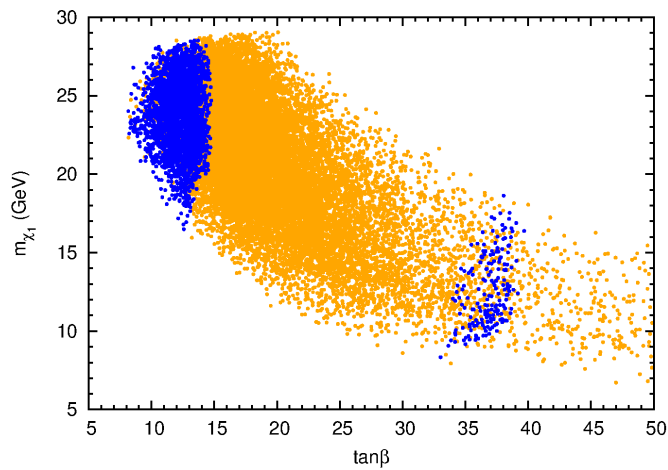}
\caption{The lightest neutralino mass as a function of $\tan\beta$. The color code is the same as in Fig.~\ref{fig:mHc-tb}. \label{fig:tb-mchi1}}
\end{figure}

The orange (light gray) points in Fig.~\ref{fig:mHc-tb} satisfy the constraints of Tab.~\ref{Tab:condition},
except the ones belonging to group (i). The green (medium gray) points give
$B\to\tau\nu$ within the experimental range, while the blue (dark gray) points satisfy all the constraints, in particular
$R_{\ell23}$. The results are consistent with the discussion of the previous section: the interplay between
$B\to\tau\nu$ and $R_{\ell23}$ is such that only either a low $\tan\beta$ region survives or a tuned 
strip for $\tan\beta\gtrsim 30$. 

In Fig.~\ref{fig:tb-mchi1} we show how this can be translated in a bound on the lightest neutralino mass.
In the figure, the lightest neutralino mass $m_{\tilde{\chi}^0_1}$ is plotted as a function of $\tan\beta$.
The plot is shaped by the upper and the lower limit on $\Omega_{\rm DM} h^{2}$ (from below and above respectively).
We see that the requirement of a relic density consistent with WMAP makes increase the neutralino mass
quite rapidly when $\tan\beta$ decreases, as previously observed in the literature, see for instance \cite{Bottino:2002ry}.
In particular light neutralinos ($\sim 10$ GeV) can be achieved only for large values of $\tan\beta$. 
As in Fig.~\ref{fig:mHc-tb}, the orange (light gray) points do not fulfill the bounds of the group (i), while 
for the blue (dark gray) all constraints are satisfied. We see that in the low $\tan\beta$ region the lightest 
neutralino mass is bounded to be $m_{\tilde{\chi}^0_1} \gtrsim 16$ GeV, while for the points
lying in the large $\tan\beta$ strip it can be as low as $\sim 8$ GeV. This last result is consistent
with the findings of Ref.~\cite{Fornengo:2010mk}, despite the fact that in that analysis $R_{\ell23}$ 
was not taken into account.
\begin{figure}[t]
\centering
\includegraphics[height=0.45\textwidth,angle=-90]{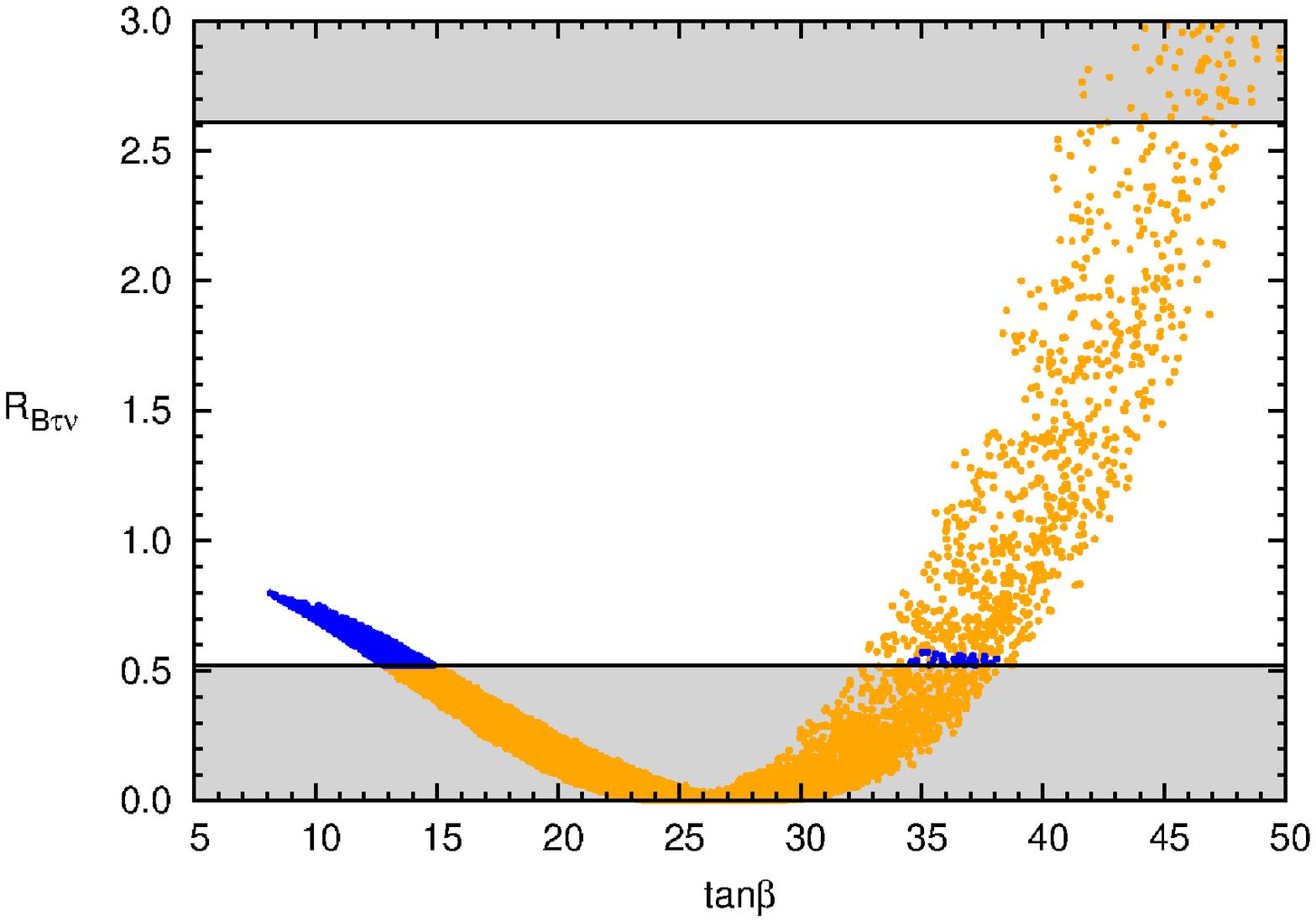}
\includegraphics[height=0.45\textwidth,angle=-90]{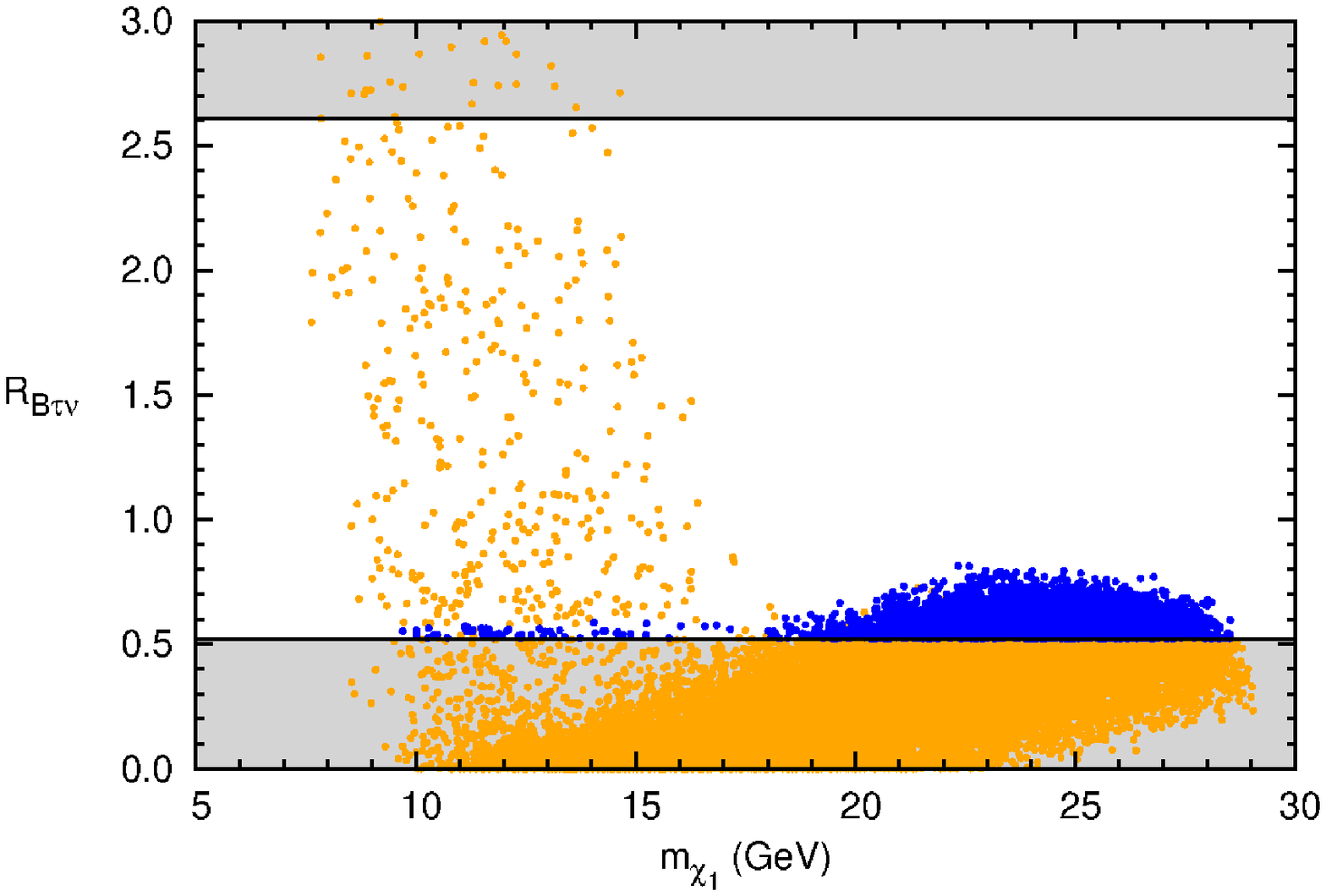}
\includegraphics[height=0.45\textwidth,angle=-90]{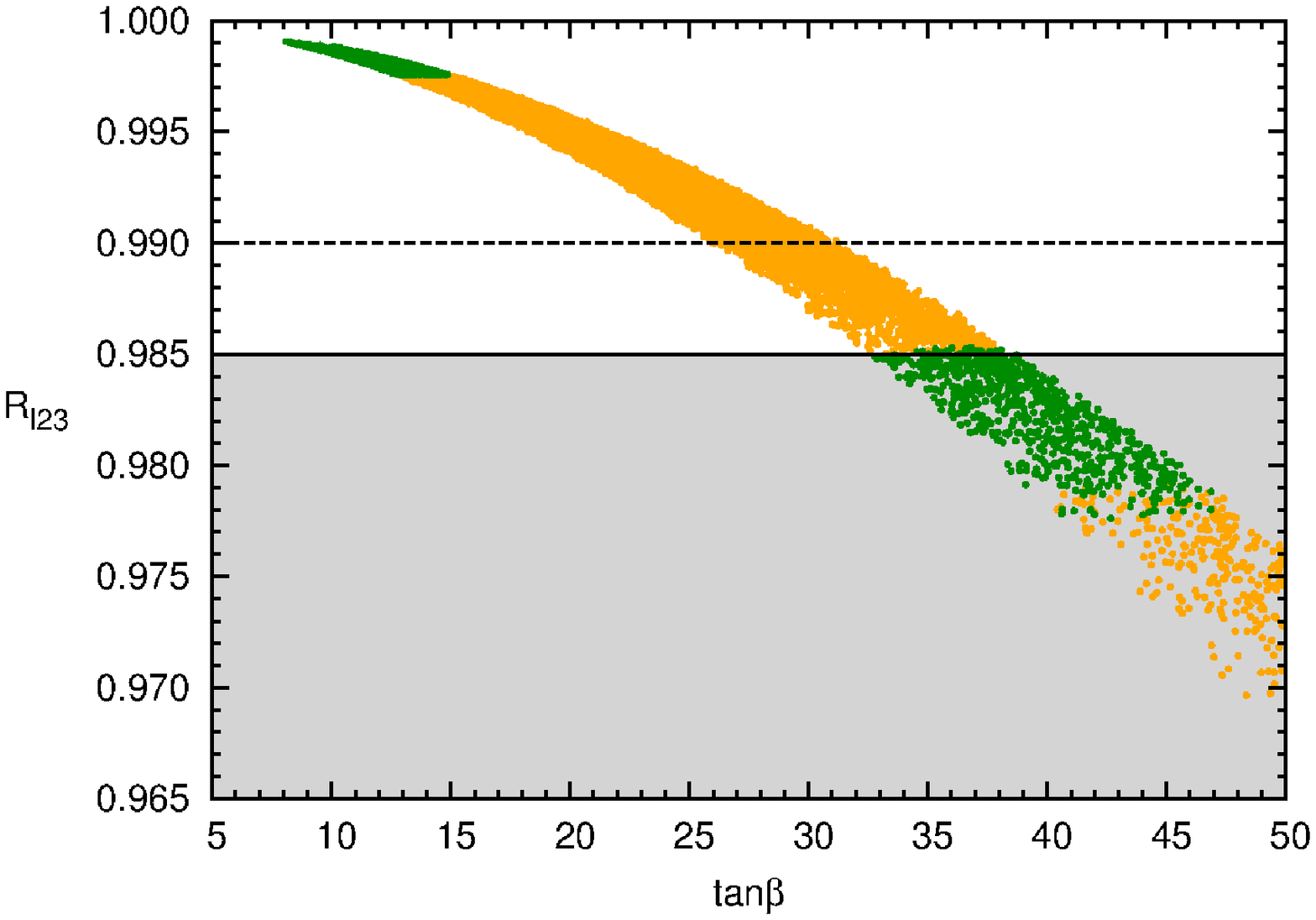}
\includegraphics[height=0.45\textwidth,angle=-90]{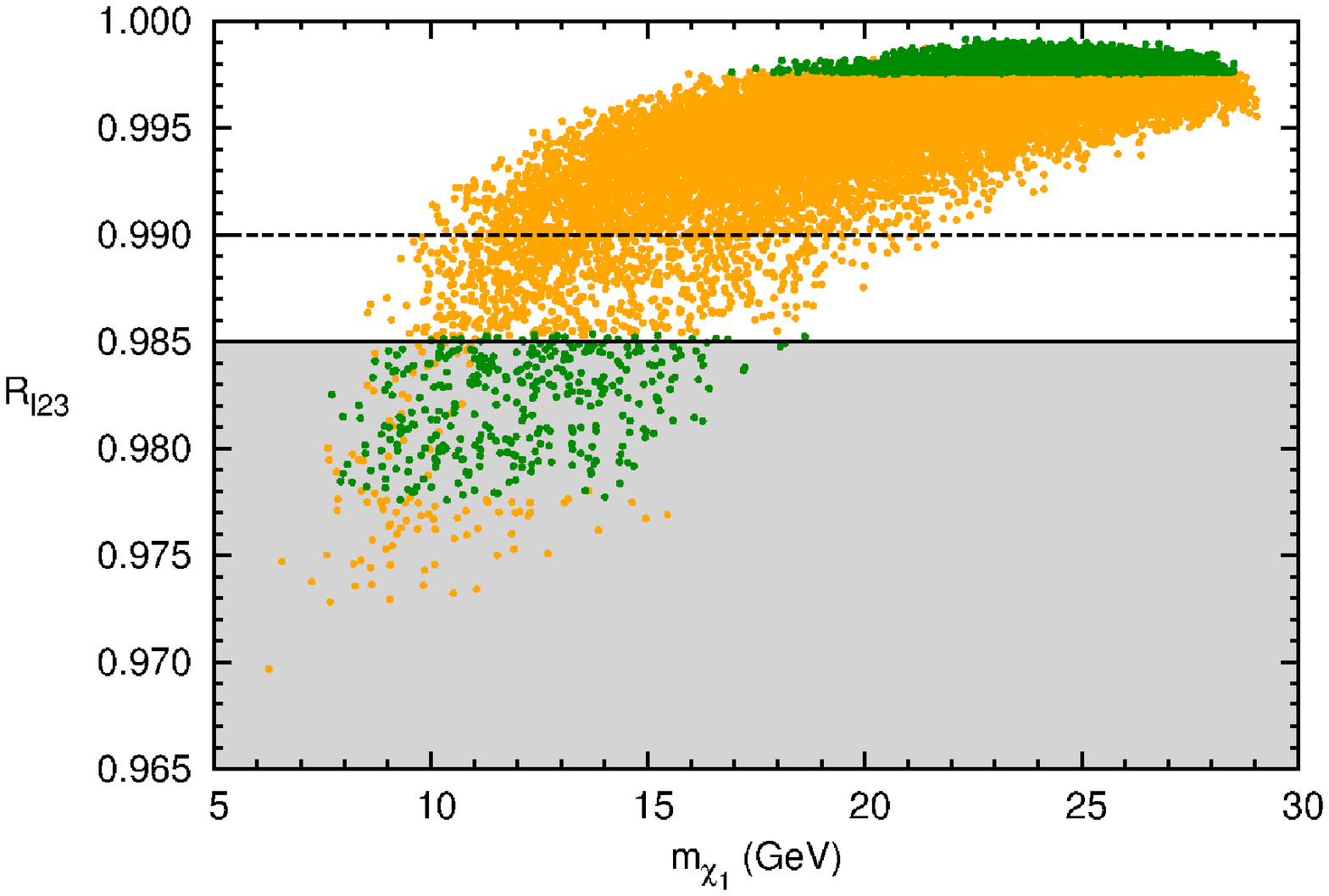}
\caption{First row: the observable $R_{B\tau\nu}$ as a function of $\tan\beta$ (left) and $m_{\tilde{\chi}^0_1}$ (right).
The shaded regions are excluded by $B\to\tau\nu$ at the 95\% C.L.. 
Blue (dark gray) points satisfy all the constraints of Tab.~\ref{Tab:condition}.
Second row: the Kaon physics observable $R_{\ell 23}$ as a function of $\tan\beta$ (left) and $m_{\tilde{\chi}^0_1}$ (right).
The shaded region is excluded by $R_{\ell 23}$ at the 95\% C.L., the dashed line represents the previous bound 
$R_{\ell23}> 0.990$ (see the text for details).
Green (medium gray) points satisfy the $B\to\tau\nu$ constraint.\label{fig:RBtn}}
\end{figure}

Given the relevance in the present discussion of the interplay between $B\to\tau\nu$ and $R_{\ell23}$,
let us have now a closer look at these observables. In the first row of Fig.~\ref{fig:RBtn}, we show the ratio
$R_{B\tau\nu}$, defined in Eq.~(\ref{eq:btn}), versus $\tan\beta$ (left panel) and $m_{\tilde{\chi}^0_1}$
(right panel). The shaded regions in the plot are excluded by $R_{B\tau\nu}$ at the 95\% C.L..
Consistently with the discussion in section~\ref{sec:lowE}, we see that 
$R_{B\tau\nu}$ decreases with increasing $\tan\beta$ (which correspond to decreasing values of $m_{\tilde{\chi}^0_1}$)
until getting outside the experimentally allowed range. Then, $R_{B\tau\nu}$ acquires vanishing small values, corresponding
to the charged Higgs contribution exactly cancelling the SM contribution. For larger values of $\tan\beta$, the 
large charged Higgs contribution can make $R_{B\tau\nu}$ to re-enter the allowed range. However, only a thin strip
of points (the blue/dark gray ones) close to the boundary are not excluded by the full set of our constraints, in particular $R_{\ell23}$.
This can better seen in the second row of Fig.~\ref{fig:RBtn}, where $R_{\ell23}$ is plotted versus $\tan\beta$ 
(left panel) and $m_{\tilde{\chi}^0_1}$.
(right panel). As expected from Eq.~(\ref{eq:rl23}), $R_{\ell23}$ decreases with $\tan\beta$ (i.e. with decreasing
$m_{\tilde{\chi}^0_1}$). The two green (medium gray) points regions are allowed by $B\to\tau\nu$. As discussed above, the one with large
$\tan\beta$ (smaller $m_{\tilde{\chi}^0_1}$) corresponds to the charged Higgs contribution overcompensating the SM one. As we can see, 
this region is almost excluded by $R_{\ell23}$, except for few surviving points with $R_{\ell23}$ about 2-$\sigma$ 
away from the experimental central value $R^{\rm cv}_{\ell23} = 0.999$~\cite{Antonelli:2010yf}. In the figure, we also show the previous lower bound of Ref.~\cite{Antonelli:2008jg}, $R_{\ell23}> 0.990$, which would have excluded completely the large $\tan\beta$ region consistent with $B\to\tau\nu$.
Needless to say, an improvement of the experimental determination of $R_{\ell23}$ or $R_{B\tau\nu}$ could completely exclude 
the large $\tan\beta$ region. This would translate in a lower bound on the lightest neutralino mass 
at least of $m_{\tilde{\chi}^0_1} \gtrsim 16$ GeV, as we can see from Fig.~\ref{fig:RBtn}.

Finally we comment about the LEP neutralino searches, 
which constrain the pair production cross-section 
$\sigma(e^{+}e^{-} \rightarrow
\tilde{\chi}^{0}_{1}\tilde{\chi}^{0}_{2,3})$~\cite{Ellis:1983er,Bartl:1986hp},
and 
the contribution to the invisible width of the $Z$ boson of the decay
$Z\to\tilde{\chi}^{0}_{1}\tilde{\chi}^{0}_{1}$~\cite{Barbieri:1987hb,Heinemeyer:2007bw}. 
The pair production process is mediated not only by selectron but also 
$Z$ boson. Therefore, even if a large value is taken for the slepton 
mass $m_{\tilde{\ell}}$, the process does not fade and 
constrains the low energy parameters.
Both the $Z$-mediated contribution of the pair production
and the invisible decay width of $Z$ boson are proportional to 
the higgsino component in the lightest neutralino.
As discussed in Sec.~\ref{Sec:relic}, the higgsino components
are controlled by $\mu$, and thus these observations require somewhat 
larger values of $\mu$.
On the other hand, such a parameter choice reduces the neutralino 
annihilation cross-section. In order to compensate this
reduction and reproduce the correct relic density, 
$M_{1}$, $\tan\beta$, and $m_{A}$ must be adjusted.
As already noticed in Ref.~\cite{Fornengo:2010mk}, 
the invisible decay width does 
not give any impact on the limit of the lightest neutralino mass, 
even considering a less conservative limit 
$\Gamma(Z\to \tilde{\chi}^{0}_{1}\tilde{\chi}^{0}_{1})< 2~{\rm MeV}$.
We also inspect the impact of the constraint from the pair production
process and find that it does not modify our lower bounds
on the lightest neutralino mass.

\section{Phenomenological consequences}
\label{sec:prospects}

\subsection{Higgs sector}
\begin{figure}[t]
\centering
\includegraphics[height=0.45\textwidth,angle=-90]{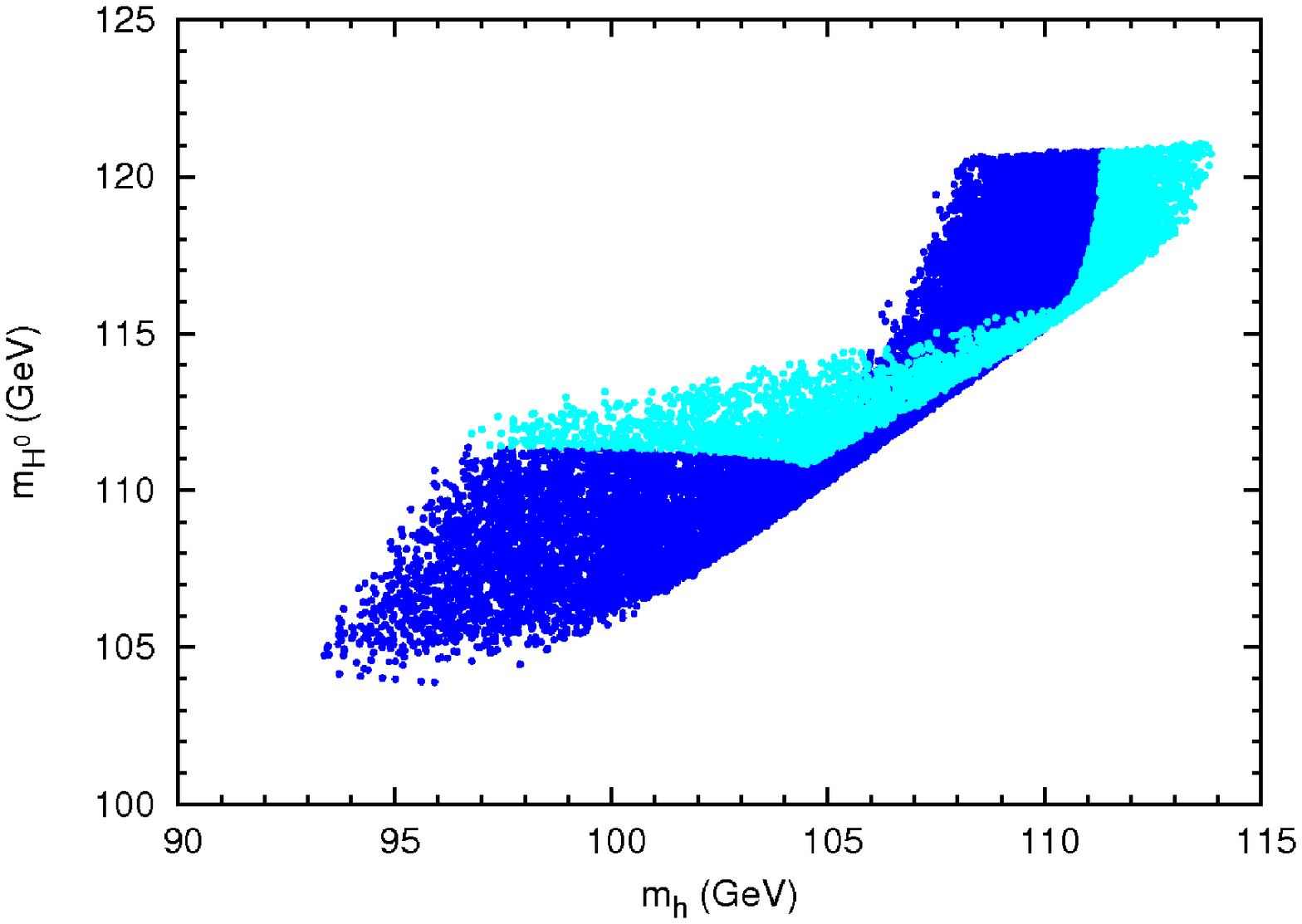}
\includegraphics[height=0.45\textwidth,angle=-90]{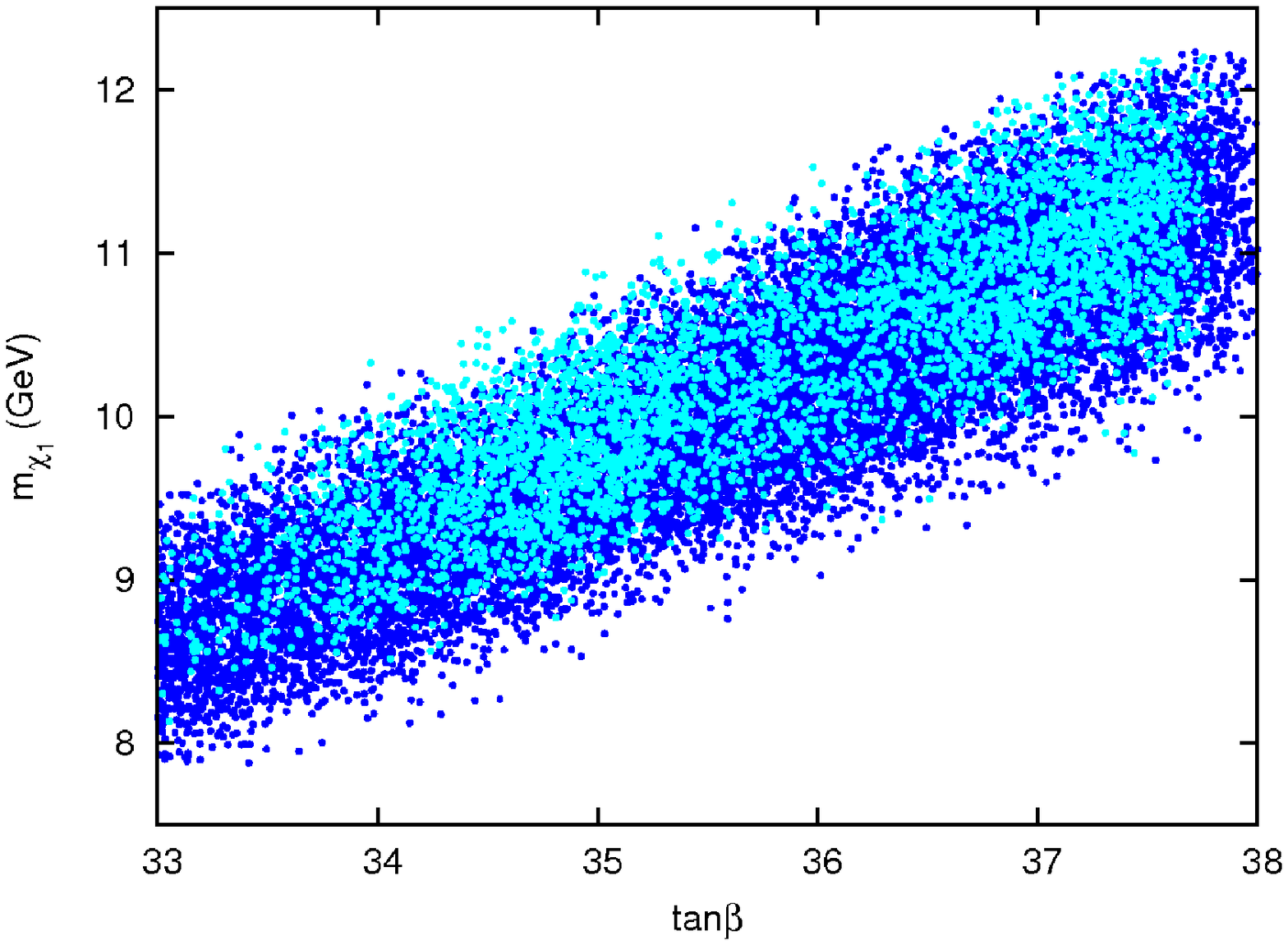}
\caption{Left: the heavy neutral Higgs mass versus the light Higgs mass for the points lying in the large $\tan\beta$ 
strip of Fig.~\ref{fig:mHc-tb} with $\mu\leq 120$ GeV. The dark blue (dark gray) points fulfill all the constraints
of Tab.~\ref{Tab:condition}, the light blue (light gray) points escape the LEP Higgs bosons search (see the text for details).
Right: the same as before in the $\tan\beta$-$m_{\tilde{\chi}^0_1}$ plane.\label{fig:LEP-Higgs}}
\end{figure}
In this section, we discuss in greater detail the predictions of the light neutralino MSSM scenario for the Higgs sector.
Let us first discuss the constraints on the masses of the Higgs bosons from LEP experiments.
As shown in Tab.~\ref{Tab:condition}, we have applied so far a conservative bound on the light Higgs mass, 
$m_h > 92.8$ GeV, which is valid in the so-called anti-decoupling regime $\sin^{2} (\alpha - \beta)\rightarrow 0$.\footnote{
We remind that $\alpha$ is the mixing angle, which links the CP-even Higgs mass eigenstates to the interaction
eigenstates $H_u^0$, $H_d^0$. For a review on the MSSM Higgs sector, we refer to Ref.~\cite{Djouadi:2005gj}.}
In such a regime, however, the LEP limit on the SM Higgs mass should be applied to the heaviest eigenstate, $H^0$.
In the decoupling regime, $\sin^{2} (\alpha - \beta)\rightarrow 1$, the lightest Higgs couplings become SM-like and the 
LEP bound on the SM Higgs should be recovered. In order to deal with intermediate regimes as well, we made use of the LEP exclusion
plot in the $m_h$-$\sin^{2} (\alpha - \beta)$ plane, published in
Ref.~\cite{Schael:2006cr}. Let us see how the light neutralino scenario
is affected. In the left panel of Fig.~\ref{fig:LEP-Higgs}, we plot the resulting values for $m_h$ and $m_{H^0}$ 
within the large $\tan\beta$ strip discussed in the previous section. 
For collecting more points, we performed a focussed scan of the large $\tan\beta$ region 
taking $\mu \leq 120$ GeV (this choice does not affect the lower
bound on the neutralino mass, since lighter neutralinos requires smaller values of $\mu$).
The light blue (light gray) points survive the light and heavy Higgs mass constraints of Ref.~\cite{Schael:2006cr}, once
3 GeV of theoretical uncertainty on the determination of the Higgs masses has been taken into account.
The results can be easily interpreted: for an heavier $H^0$ ($m_{H^0}\gtrsim 115$ GeV), the decoupling regime is approached, the light
Higgs gets SM-like and the points with  $m_{h}< 111$ GeV are excluded. On the other hand, when $H^0$ gets light, since $\tan\beta$
is large, we are in the anti-decoupling regime where $H^0$ takes the
role of the SM Higgs boson. 
As a consequence, the points with $m_{H^0}< 111$ GeV are excluded too.
In the right panel of Fig.~\ref{fig:LEP-Higgs}, we plot the same points in the $\tan\beta$-$m_{\tilde{\chi}^0_1}$ plane.  
Even though the density of points is consistently reduced by the LEP limit on Higgs masses, we see no relevant modification of 
the lower bound on the lightest neutralino mass, $m_{\tilde{\chi}^0_1}\gtrsim 8$ GeV, we found in the large $\tan\beta$ strip.
Still, it is interesting to notice that a quite light Higgs spectrum is predicted within the strip.
We checked that the low $\tan\beta$ region is not significantly affected by the Higgs mass bounds neither.
\begin{figure}[t]
\centering
\includegraphics[width=0.55\textwidth]{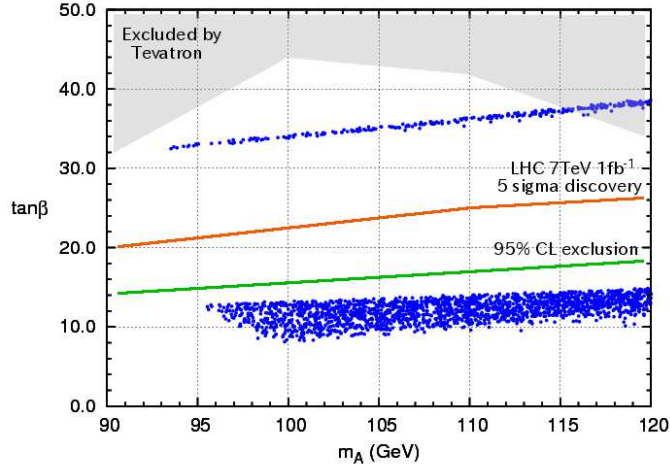}
\caption{Constraint from the Higgs boson search at Tevatron (gray shade)
and the expected discovery (above the orange curve) and exclusion (above the green curve) potential
of LHC at end of 2011 (with 7 TeV of center-of-mass energy  
and 1 fb$^{-1}$ of integrated luminosity) \cite{cms}.
Parameter points which satisfy all the constraints listed 
in Tab.~\ref{Tab:condition} are shown together with blue dots
(cf. Fig.~\ref{fig:mHc-tb}).
The large $\tan \beta$ strip will be tested soon.
\label{fig:LHC-Higgs}}
\end{figure}

A light Higgs sector as predicted by the MSSM with light neutralino has very good prospects of being tested soon at the LHC. 
Recently, CMS has published the expected sensitivity of the Higgs bosons search in the channel $pp\to bb\,\Phi \to bb\, \tau\tau$
(where $\Phi = A,\,H^0$), with 7 TeV of center of mass energy and 1 $\rm fb^{-1}$ of integrated luminosity \cite{cms}. The LHC
potential of testing the Higgs sector of our scenario is depicted in Fig.~\ref{fig:LHC-Higgs}, where we replot the two regions
which evade all the constraints in the $m_A$-$\tan\beta$ plane. Above the orange line a 5-$\sigma$ discovery is possible,
while the region above the green line can be excluded at 95 \% C.L. \cite{cms}. We also show the Tevatron exclusion \cite{tevatron} 
as a gray shaded region.
We see that the large $\tan\beta$ strip is excluded by Tevatron experiments only for larger values
of $m_A$ and $\tan\beta$, which correspond to larger $m_{\tilde{\chi}^0_1}$, such that the lower bound on the 
light neutralino mass is not affected.
Remarkably, the MSSM region corresponding to lightest neutralino masses seems to be accessible at the LHC in the near future.
In fact, wee see from Fig.~\ref{fig:LHC-Higgs} that the large $\tan\beta$ strip should be completely tested by the present LHC run,
i.e. either discovered or fully excluded with the data collected by end of 2011. 
On the contrary, more luminosity should be needed to probe the low $\tan\beta$ region, corresponding to $m_{\tilde{\chi}^0_1}\gtrsim 16$ GeV. 

\subsection{Low-energy observables}
\begin{figure}[t]
\centering
\includegraphics[height=0.45\textwidth,angle=-90]{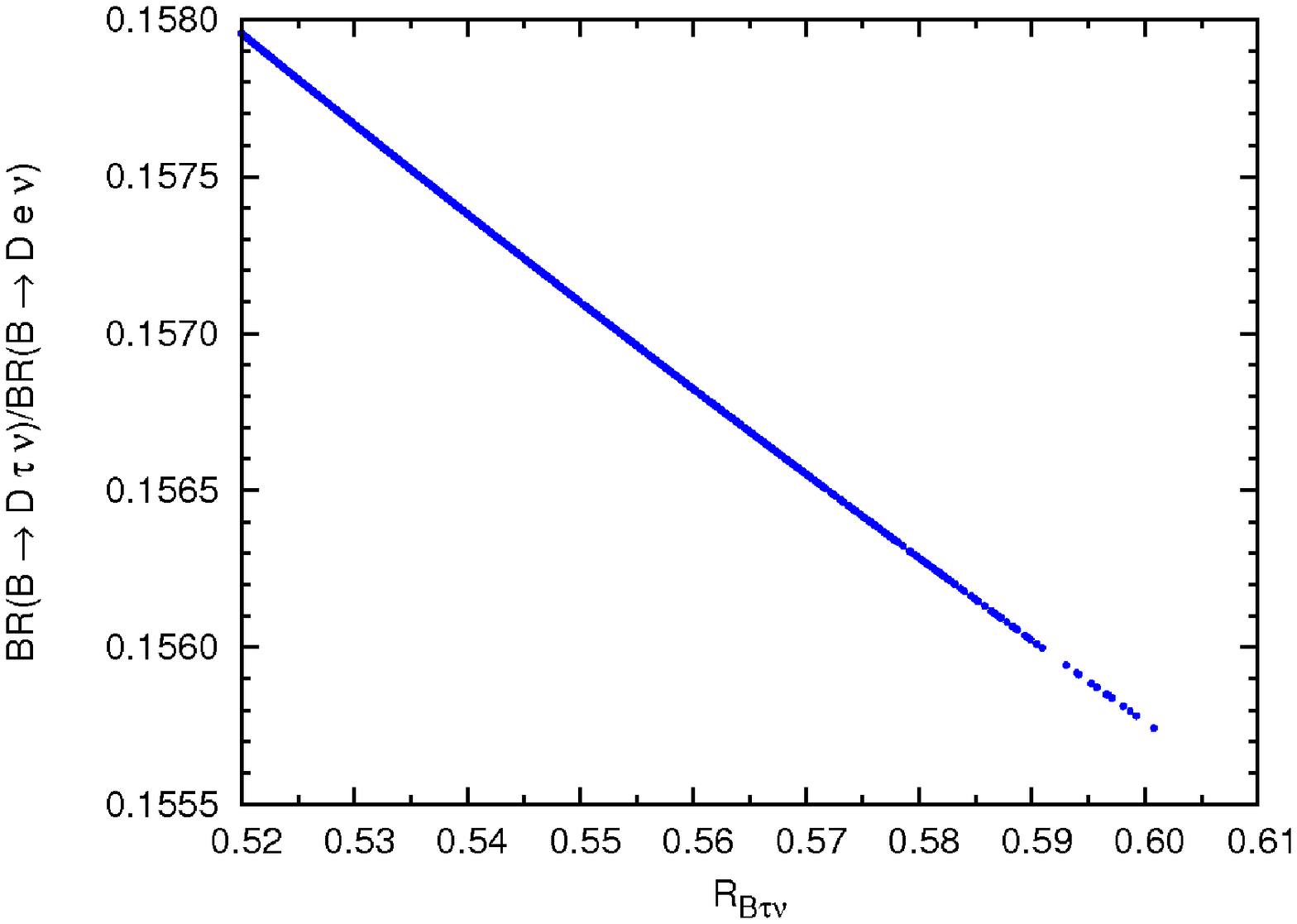}
\includegraphics[height=0.45\textwidth,angle=-90]{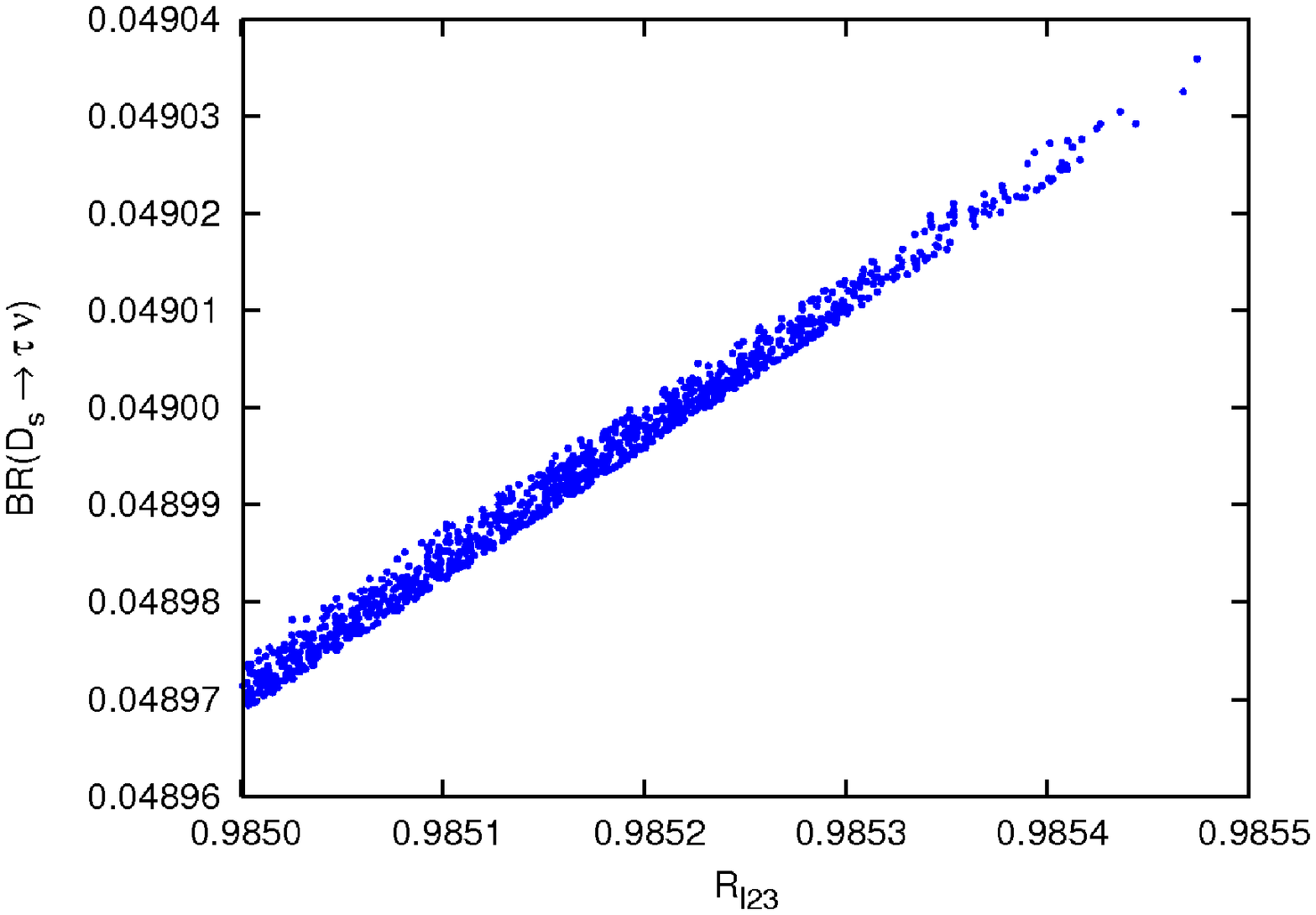}
\caption{Predicted ranges for group (i) observables within the large $\tan\beta$ strip of Fig.~\ref{fig:mHc-tb}.
Left: ${{\rm BR}(B \rightarrow D \tau \nu)}/{{\rm BR}(B \rightarrow D e \nu)}$ versus $R_{B\tau\nu}$.
Right: ${\rm BR}(D_s \rightarrow \tau \nu)$ versus $R_{\ell23}$.
\label{fig:lowE-prospects}}
\end{figure}

Let us now concentrate on the predictions for the low-energy observables of the parameter space points lying in the large 
$\tan\beta$ strip. As stressed in the previous section, this strip requires a fine tuning of the parameters, but it cannot
be excluded by the present low-energy data. However, it predicts several observables to deviate from the current central values
almost at the 2-$\sigma$ level. More interestingly, the predictions for the observables lie in very narrow ranges, as a consequence
of the complementarity of the $B\to\tau\nu$ and $R_{\ell23}$ bounds. This is shown in Fig.~\ref{fig:lowE-prospects}, where the
predictions for the group (i) observables are depicted, 
${{\rm BR}(B \rightarrow D \tau \nu)}/{{\rm BR}(B \rightarrow D e \nu)}$ versus $R_{B\tau\nu}$ on the left, 
${\rm BR}(D_s \rightarrow \tau \nu)$ versus $R_{\ell23}$ on the right. We notice the striking correlations
among these observables, as expected by the similar dependence on $\tan\beta$ and $m_{H^\pm}$ they manifest (cf.
Eqs.~(\ref{eq:btn}, \ref{eq:rl23}, \ref{eq:xdln}, \ref{eq:dtn})), and the very limited ranges the bounds on 
$B\to\tau\nu$ and $R_{\ell23}$ allow. Comparing that with the experimental values reported in Tab.~\ref{Tab:condition},
we see that this region of the parameter space is in tension with the experiments between 1 and 2 $\sigma$
for all the group (i) processes. We come again to the conclusion that the large $\tan\beta$ region seems to be quite unlikely and that, 
more importantly, it might be fully probed by the improvement of the experimental determination of any of the observables
shown in Fig.~\ref{fig:lowE-prospects}.

\subsection{SUSY parameter space}
\begin{figure}[t]
\centering
\includegraphics[height=0.45\textwidth,angle=-90]{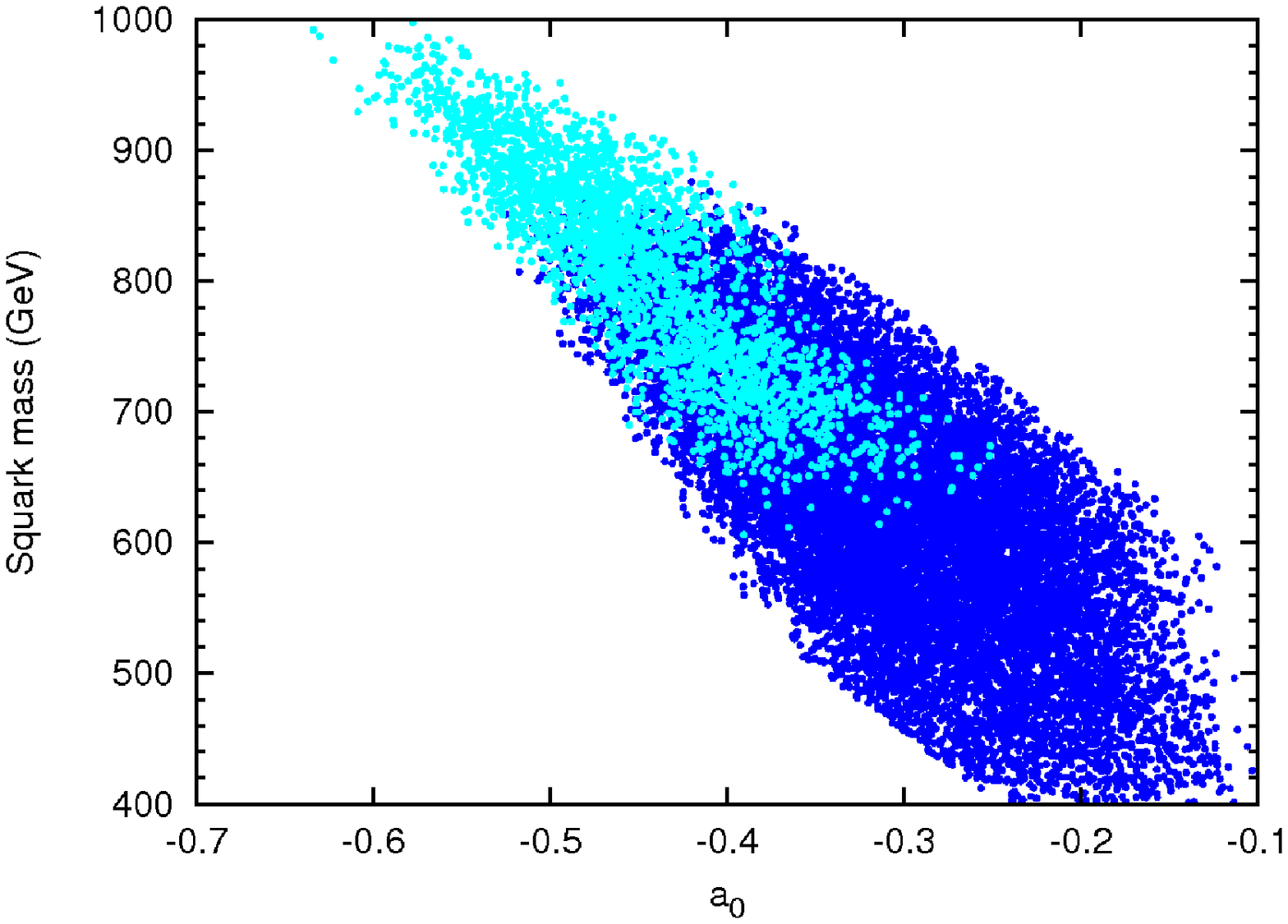}
\includegraphics[height=0.45\textwidth,angle=-90]{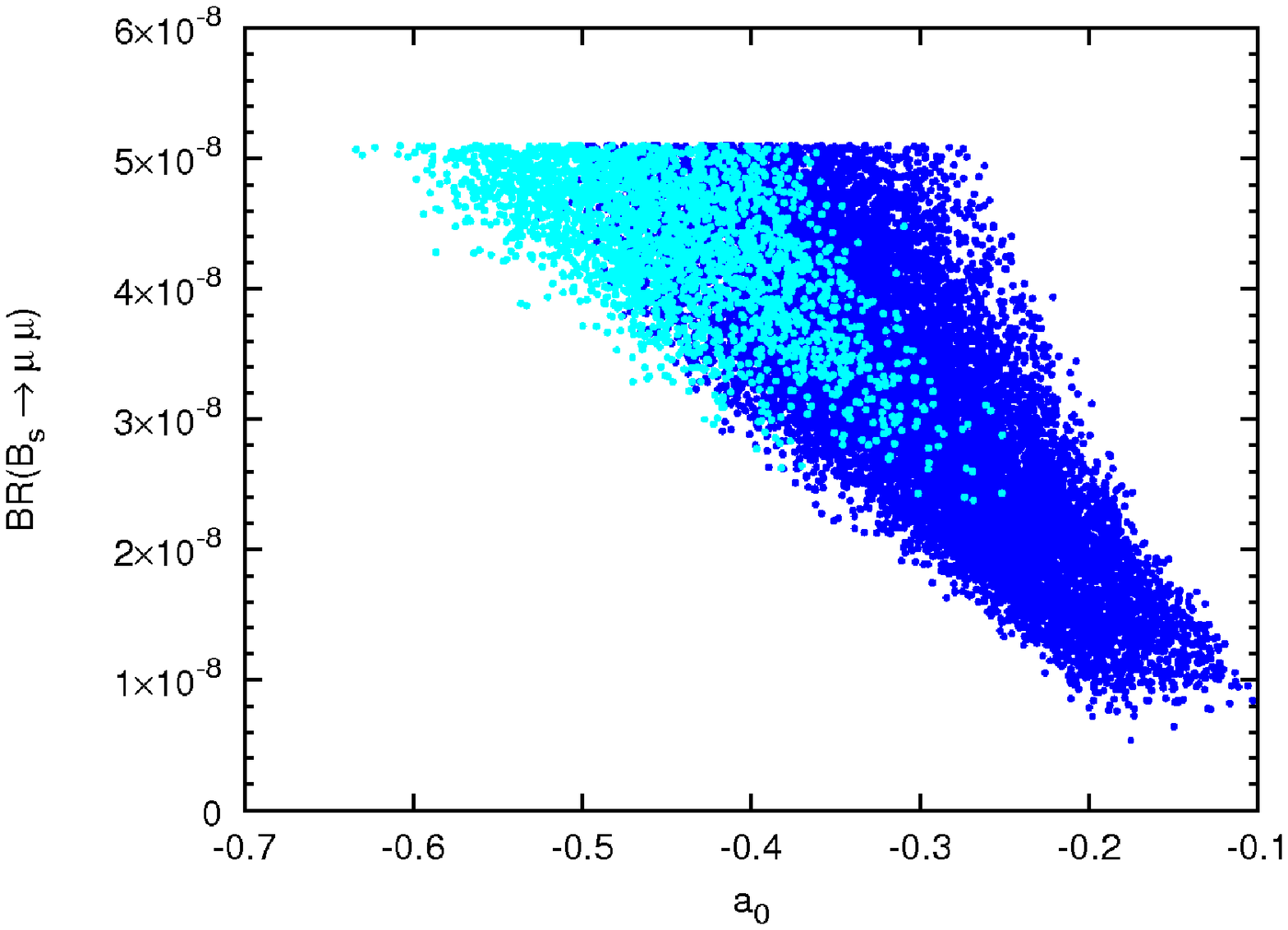}
\caption{Left: the points lying in the large $\tan\beta$ 
strip of Fig.~\ref{fig:mHc-tb} with $\mu\leq 120$ GeV in the $a_0$-$m_{\tilde q}$ plane. Light blue (light gray) points satisfy
the LEP bounds on Higgs bosons masses. Right: BR($B_s\to \mu\mu$) as a function of $a_0$ for the same sample of points.
\label{fig:a0-msq}}
\end{figure}

Let us now have a closer look at the SUSY parameter space, which is selected by the requirement of a light neutralino MSSM fulfilling 
all the imposed experimental constraints. As we have already discussed, the constraints of group (i) have only a mild dependence on SUSY
parameters through SUSY threshold corrections. On the contrary, the observables of group (ii) and the Higgs bosons masses depend considerably on SUSY parameters.

Out of the two regions in Fig.~\ref{fig:mHc-tb} that fulfill the constraints of Tab.~\ref{Tab:condition}, 
let us first consider the low $\tan\beta$ region. We find that the $b\to s\gamma$ bound requires $m_{\tilde q} \lesssim 800$ GeV
and $a_0 \lesssim -0.4$. This is a consequence of the necessary compensation between the charged Higgs and the chargino contributions
to $b\to s\gamma$. In fact, $m_{\tilde q}$ has to be rather light and $|a_0|$ to be sizeable such that the chargino contribution is large
enough to compensate the charged Higgs contribution, which is large due to the light $H^\pm$. The sign of $a_0$ is fixed by the 
requirement that the two contributions have opposite signs, i.e. $\mu A_t < 0$. Of course, considering heavier neutralinos 
(our scan is limited to $m_{\tilde{\chi}^0_1}\lesssim 30$ GeV) would allow heavier $H^\pm$ and therefore heavier squarks.
Imposing the Higgs mass bound discussed in the previous subsection, we get in addition $m_{\tilde q} \gtrsim 500$ GeV
and $a_0 \lesssim -0.8$, since a sizeable one-loop correction to the light Higgs mass is required to evade the LEP limit.
Let us stress that, for simplicity, we considered degenerate squarks and the bounds mentioned above only apply to the stop masses.
In fact, both $b\to s\gamma$ and the Higgs mass are only sensitive to the stop mass. We checked that, after 
relaxing the squark degeneracy, we get no constraint on the first generations squark masses. Consequently the results of the recent 
and near future SUSY searches at the LHC \cite{Khachatryan:2011tk,daCosta:2011qk} are not directly applicable to our case \cite{Scopel:2011qt}.

In the second allowed region of Fig.~\ref{fig:mHc-tb}, i.e. the large $\tan\beta$ strip, the $b\to s\gamma$ and $B_s\to \mu\mu$
bounds induce a stringent correlation between $a_0$ and $m_{\tilde q}$, as shown in the left panel of Fig.~\ref{fig:a0-msq}. 
Moreover, we see that the squark mass is bounded
to be $m_{\tilde q} \lesssim 1000$ GeV and $-0.6 \lesssim a_0 \lesssim -0.1$. The squarks can be heavier than in the low
$\tan\beta$ regime, since the chargino contribution grows with
$\tan\beta$. The value of $|a_0|$ cannot be too large, due to 
the $B_s\to \mu\mu$ constraint, which is particularly strong in the large $\tan\beta$ regime and grows proportional to $A_t^2$. 
As a result, the plot is bounded from below by the limit on $B_s\to \mu\mu$. 
We also notice that the points which evade the LEP Higgs searches (the light blue/light gray ones in Fig.~\ref{fig:a0-msq}) require
$m_{\tilde q} \gtrsim 600$ GeV and $a_0 \lesssim -0.2$, again to have a sizeable one-loop correction to the Higgs mass.
Also in this case, the resulting range for $m_{\tilde q}$ (600$\div$1000 GeV) is only valid for the stop, 
if squarks are non-degenerate.

In the right panel of Fig.~\ref{fig:a0-msq}, we plot the resulting BR($B_s\to \mu\mu$) versus $a_0$, for the same scan of the parameters
of the left panel. From the figure, we can see how $B_s\to \mu\mu$ can impose a limit on $|a_0|$. In particular, we
are considering the most recent published 95\% C.L. limit ${\rm BR}(B_s\to \mu\mu) < 5.1\times 10^{-8}$, obtained by the D0 collaboration
using 6 ${\rm fb}^{-1}$ \cite{Abazov:2010fs}. 
A more stringent limit ${\rm BR}(B_s\to \mu\mu) < 4.3\times 10^{-8}$ has been reported by the CDF collaboration \cite{cdf}, as
a preliminary result obtained using 3.7 ${\rm fb}^{-1}$. From Fig.~\ref{fig:a0-msq}, we see that such a result
would imply $a_0 \gtrsim -0.5$ and further lower the upper bound on $m_{\tilde q}$ below the TeV level. 
As mentioned above, the limit from the Higgs bosons searches at LEP and $B_s\to \mu\mu$ are complementary in constraining $a_0$. 
As a consequence, a lower bound for the possible value of BR($B_s\to \mu\mu$) within the large $\tan\beta$ strip results 
from the requirement $a_0 \lesssim -0.2$ imposed by the Higgs searches. In fact, we see that the light blue (light gray) points
in Fig.~\ref{fig:a0-msq} correspond to ${\rm BR}(B_s\to \mu\mu) \gtrsim 2\times 10^{-8}$, a value which will be probed 
by LHCb in the upcoming months \cite{Aaij:2011rj}. We can conclude that $B_s\to \mu\mu$ searches represent a further 
handle to fully test the large $\tan\beta$ parameter space of the MSSM providing a light
neutralino in the near future.

In the low $\tan\beta$ region on the contrary, BR($B_s\to \mu\mu$) is mostly at the level of the SM prediction, even if
there are some points of the sample for which a sizeable deviation from the SM is possible up to  
${\rm BR}(B_s\to \mu\mu) \sim 8\times 10^{-9}$.

The slepton mass $m_{\tilde \ell}$ has a negligible impact on the observables considered so far, so that it cannot be
constrained. We can however consider the anomalous magnetic moment of the muon, $a_\mu \equiv (g-2)_\mu/2$, to get
information on $m_{\tilde \ell}$. Currently, the SM prediction differs from the experimental determination
of $a_{\mu}$ by more than 3 sigma (see e.g. \cite{Jegerlehner:2011ti}),
so that a positive new physics contribution to $a_\mu$ would be welcome to relax such a tension.
The SUSY contribution to $a_\mu$ is known to be potentially large and to depend mainly
on the slepton and chargino masses, as well as on $\tan\beta$ (with $\Delta a^{\rm SUSY}_\mu \sim \tan^2\beta$). Since
$\tan\beta$ and the chargino mass are quite well constrained by the observables we considered, 
the experimentally allowed range for $\Delta a^{\rm SUSY}_\mu$ will turn in a preferred range for $m_{\tilde \ell}$. 
Requiring $\Delta a^{\rm SUSY}_\mu$ to result within the 2-$\sigma$ range $[109,\, 407]\times 10^{-11}$ \cite{Jegerlehner:2011ti}, 
we find for the points lying in the large $\tan\beta$ strip: $250~{\rm GeV} \lesssim m_{\tilde \ell} \lesssim 1~{\rm TeV}$,
where values of the mass below the lower (above the upper) limit correspond to a too large 
(too small) SUSY contribution to $a_\mu$. In the case of the low $\tan\beta$ region, 
a contribution large enough to lower the tension below the 2-$\sigma$ level clearly requires lighter sleptons. 
In fact, we find $m_{\tilde \ell} \lesssim 630~{\rm GeV}$.

\section{Direct searches experiments}
\label{sec:direct}
\begin{figure}[t]
\centering
\includegraphics[height=0.55\textwidth,angle=-90]{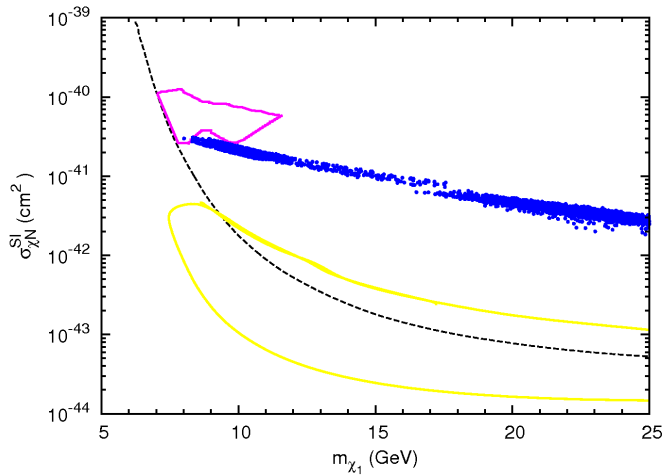}
\caption{Neutralino-nucleon scattering cross section as a function of the lightest neutralino mass for points
which satisfy all the constraints of Tab.~\ref{Tab:condition}. The pink area is the region favored by CoGeNT, the 
yellow region corresponds to the two CDMS candidates, the dashed line is the 90\% C.L. exclusion reported by XENON10. 
\label{fig:cross-section}}
\end{figure}

As mentioned in the introduction, the main motivation for studying a light neutralino scenario is provided by the recent
unexplained signals reported by some direct searches experiments.
We investigate in this section the elastic scattering cross-section of the neutralino with nucleons for the points of the 
parameter space that provide a light neutralino and satisfy the above discussed constraints. We compute the spin-independent
scattering cross-section, $\sigma^{\rm SI}_{\chi N}$, by means of the {\tt micrOMEGAs} routine \cite{micro}. 
The result is displayed in Fig.~\ref{fig:cross-section} as a function of the neutralino mass. For a comparison,
we show also in pink the region favored by CoGeNT \cite{Aalseth:2010vx}, 
in yellow the region corresponding to the two CDMS candidates \cite{Ahmed:2009zw}. 
The dashed line is the 90\% C.L. exclusion reported by XENON10 \cite{Angle:2009xb}.

In interpreting this result, one should keep in mind that several sources of uncertainties affect the computation of the
scattering cross-section on one side and, on the other side, the translation of the experimental results into favored 
or excluded regions in the WIMP mass, scattering cross-section plane. 
In particular, the uncertainty related to hadronic matrix elements enters the computation of the neutralino-nucleon scattering 
such that the resulting cross-section can be shifted to about one order of magnitude larger or smaller values \cite{Fornengo:2010mk}.  
In Fig.~\ref{fig:cross-section}, the default {\tt micrOMEGAs} values for the matrix elements are used.
On the other hand, the constraints from direct searches experiments can be relaxed by a factor 3 in the value of the cross-section 
by the uncertainties on the local DM density and velocity \cite{Belanger:2010cd}.

Taking that into account, we notice that the result shown in Fig.~\ref{fig:cross-section} is consistent with the results of \cite{Fornengo:2010mk}.
We see that, out of the two regions satisfying all the constraints, the one with 
$m_{\tilde{\chi}^0_1}\sim 10$ GeV gives the largest scattering cross-section, since it corresponds to the (tuned) large $\tan\beta$
strip in Fig.~\ref{fig:mHc-tb} and $\sigma^{\rm SI}_{\chi N}$ grows with $\tan\beta$. These points are in potential agreement with the 
regions favored by CoGeNT, CRESST and DAMA (whose viability has been recently questioned by CDMS-II \cite{Ahmed:2010wy}).
On the other hand, the low $\tan\beta$ region corresponding to $m_{\tilde{\chi}^0_1}\gtrsim 16$ GeV gives a neutralino mass range which 
might be too large for CoGeNT and CRESST, and a too small $\sigma^{\rm SI}_{\chi N}$ as well. Hence, a confirmation of the 
hints reported by these experiments would tend to disfavor it.
Still, as noticed in \cite{Bottino:2008mf, Fornengo:2010mk}, this 
region with a slightly heavier neutralino can be consistent with the annual modulation reported by DAMA 
and with the two CDMS candidate events.
However, taking into account the uncertainties, it seems to be on the edge of the exclusion 
limit provided by the negative result of XENON10 and XENON100 \cite{Angle:2009xb,Aprile:2010um,Aprile:2011hx}.
As mentioned above, the experiments are not completely in agreement with each other and the uncertainties are still too 
large to clarify the situation. Still, it seems likely that the possibility of a light neutralino in the MSSM will be fully tested 
in the next few years by direct DM searches as well.

\section{Summary and conclusion}
\label{sec:summary}
In this paper, we have analyzed the MSSM parameter space providing a light neutralino ($\sim 10\div 20$ GeV) and satisfying
all the relevant collider and flavor constraints, as well as the WMAP bound on the neutralino relic density.

We found that the request for $\Omega_{\rm DM} h^2 < 0.123$ combined with the constraints from low-energy observable which have no 
strong dependence on the SUSY parameters (but only on the Higgs sector parameters), in particular 
$B\to \tau\nu$ and $K\to\mu\nu$ ($R_{\ell 23}$), select two regions of the parameter space:
\begin{itemize}
 \item[I.]  A thin fine-tuned strip with a large value of $\tan\beta$ ($\gtrsim 30$) and $m_A \sim 100$ GeV, 
            where the neutralino mass can be as light as 8 GeV and the direct detection cross-section
             $\sigma^{\rm SI}_{\chi N} \gtrsim \mathcal{O}(10^{-41})~{\rm cm^2}$, so that it might be compatible with the exceeding
             events reported by CoGeNT and CRESST, and the DAMA annual modulation.
 \item[II.]  A low $\tan\beta$ ($\lesssim 15$) region, where the charged Higgs contribution to $B\to \tau\nu$ is not required
            to overcompensate the SM contribution. This region gives $m_{\tilde{\chi}^0_1}\gtrsim 16$ GeV 
             and $\sigma^{\rm SI}_{\chi N} \sim 10^{-42}\div 10^{-41}~{\rm cm^2}$.
\end{itemize}
Within region I, which is a strip that barely survives the competing constraints from $B\to\tau \nu$ and $R_{\ell 23}$, several low-energy
observables deviate from the SM predictions and from the experimental central values almost at the 2-$\sigma$ level.
This is the case of the SUSY independent observables of group (i), which then represent a crucial handle to probe this scenario.
Moreover, region I provides ${\rm BR}(B_s\to \mu\mu) \gtrsim 2\times 10^{-8}$, a value in the reach of the LHCb 
experiment in the upcoming months. Finally, heavy Higgs searches at the LHC should also probe this large $\tan\beta$ strip with
1 ${\rm fb}^{-1}$ of collected data, i.e. by end of 2011. 
Even though it might already appear unlikely, this set-up of the MSSM parameters corresponding to a light neutralino ($\sim 10$ GeV) 
is not excluded at present. On the other hand, we can conclude that it will be fully tested in the near future 
in several independent ways. 

Region II should escape such early experimental searches and require more years of data taking. Clearly, 
it would be the only possible set-up left within the MSSM providing a quite light neutralino ($m_{\tilde{\chi}^0_1}\gtrsim 16$ GeV),
in case of a negative result of the experimental tests for region I.
The most promising way of probing region II seems to rely on direct searches experiments. 
Even though the situation is not well established yet (region II seems to be disfavored by XENON, but compatible with DAMA annual 
modulation and the two CDMS candidates), there are good perspectives for the experiments currently taking data (such as XENON100)
to completely probe that region and, more in general, the light neutralino MSSM scenario.

Finally, we make a few comments on the recent studies~\cite{Kuflik:2010ah,Feldman:2010ke,Vasquez:2010ru,Fornengo:2010mk}.
After taking into account the constraints considered in these works as well as $R_{\ell23}$ and the LEP Higgs bounds
in the way explained in section \ref{sec:prospects},
we find a lower limit for the neutralino mass, $m_{\tilde{\chi}^0_1}\gtrsim 8$ GeV, which is consistent with the results of
Ref.~\cite{Fornengo:2010mk}. However, the points giving such light neutralinos correspond to a region that survives 
on a careful balance among the constraints (especially $B\to\tau\nu$ and
$R_{\ell23}$) and consequently requires a tuning of the parameters. 
Therefore, if one considers slightly more stringent allowed ranges for the constraints, 
this region might disappear, as it seems to be the case of Ref.~\cite{Vasquez:2010ru}.
Hence, the lower limit, $m_{\tilde{\chi}^0_1}\gtrsim 16$ GeV, of the low $\tan\beta$ region seems to be more robust.

\begin{acknowledgments}
 We are grateful to  G.~B{\'e}langer, A.~J.~Buras, G.~Isidori, J.~Jones-Perez, F.~Mahmoudi and P.~Paradisi 
 for useful discussions or communications, and to J.~Redondo for his continuous moral support and encouragement.
 One of us (Y.~T.) wishes to thank Max-Planck-Institute f\"ur Physik for very warm hospitality 
 during the preparation of this work.
\end{acknowledgments}

\section*{Note Added}
After the completion of this work, the CMS collaboration released the first results of the search for neutral Higgs bosons 
in the channel $pp\to X\,\Phi \to \tau\tau$, based on an integrated luminosity of 36 $\rm pb^{-1}$ \cite{Chatrchyan:2011nx}.
Since no excess has been observed, they pose an upper bound on $\tan\beta$ ($\tan\beta \lesssim 29$ for $m_A = 100$ GeV), which, 
even taking into account theoretical uncertainties, disfavors the large $\tan\beta$ strip discussed in this paper (see however \cite{baglio}). 
If this result will be confirmed, this would represent the first test excluding the large $\tan\beta$ region,
as we discussed in section \ref{sec:prospects}. In this case, the only low $\tan\beta$ region would remain viable, corresponding
to a lower bound on the lightest neutralino mass of $m_{\tilde{\chi}^0_1}\gtrsim 16$ GeV.


%

\begin{thebibliography}{99}
%
\bibitem{Dunkley:2008ie}
  J.~Dunkley {\it et al.} [WMAP Collaboration],
  Astrophys.\ J.\ Suppl.\  {\bf 180 } (2009)  306 [arXiv:0803.0586 [astro-ph]].
%
\bibitem{Komatsu:2010fb}
  E.~Komatsu {\it et al.} [WMAP Collaboration],
  Astrophys.\ J.\ Suppl.\  {\bf 192 } (2011)  18 [arXiv:1001.4538 [astro-ph.CO]].
%
%
\bibitem{Kopp:2009qt}
  J.~Kopp, T.~Schwetz, J.~Zupan,
  JCAP {\bf 1002 } (2010)  014 [arXiv:0912.4264 [hep-ph]].
\bibitem{Regis:2010ay}
M.~Regis,
 arXiv:1008.0506 [hep-ph].
\bibitem{Schwetz:2010gv}
  T.~Schwetz,
  arXiv:1011.5432 [hep-ph].



\bibitem{Bernabei:2008yi}
  R.~Bernabei {\it et al.} [DAMA Collaboration],
  Eur.\ Phys.\ J.\  {\bf C56 } (2008)  333  [arXiv:0804.2741 [astro-ph]].
%
\bibitem{Bernabei:2010mq}
  R.~Bernabei, P.~Belli, F.~Cappella, R.~Cerulli, C.~J.~Dai, A.~d'Angelo, H.~L.~He, A.~Incicchitti {\it et al.},
  Eur.\ Phys.\ J.\  {\bf C67 } (2010)  39 [arXiv:1002.1028 [astro-ph.GA]].
%
\bibitem{Aalseth:2010vx}
  C.~E.~Aalseth {\it et al.} [CoGeNT Collaboration],
  [arXiv:1002.4703 [astro-ph.CO]].
%
\bibitem{NewCRESST:2011}
J.~Schmaler [CRESST Collaboration], ``Results from the CRESST Dark Matter Search'', 
Talk given at DPG, 28. March - 1. April 2011, Karlsruhe, Germany.
%
\bibitem{Hooper:2010uy}
  D.~Hooper, J.~I.~Collar, J.~Hall, D.~McKinsey,
  Phys.\ Rev.\  {\bf D82 } (2010)  123509
  [arXiv:1007.1005 [hep-ph]].
%
\bibitem{Kuflik:2010ah}
  E.~Kuflik, A.~Pierce, K.~M.~Zurek,
  Phys.\ Rev.\  {\bf D81 } (2010)  111701 [arXiv:1003.0682 [hep-ph]].
%
\bibitem{Feldman:2010ke}
  D.~Feldman, Z.~Liu, P.~Nath,
  Phys.\ Rev.\  {\bf D81 } (2010)  117701 [arXiv:1003.0437 [hep-ph]].
%
\bibitem{Vasquez:2010ru}
  D.~A.~Vasquez, G.~B{\'e}langer, C.~Boehm, A.~Pukhov, J.~Silk,
  Phys.\ Rev.\  {\bf D82 } (2010)  115027 [arXiv:1009.4380 [hep-ph]].
%
\bibitem{Fornengo:2010mk}
  N.~Fornengo, S.~Scopel, A.~Bottino,
  Phys.\ Rev.\  {\bf D83 } (2011)  015001 [arXiv:1011.4743 [hep-ph]].

\bibitem{Hooper:2002nq}
  D.~Hooper, T.~Plehn,
  Phys.\ Lett.\  {\bf B562} (2003)  18
  [hep-ph/0212226].
%
\bibitem{Dreiner:2009ic}
  H.~K.~Dreiner, S.~Heinemeyer, O.~Kittel, U.~Langenfeld, A.~M.~Weber, G.~Weiglein,
  Eur.\ Phys.\ J.\  {\bf C62 } (2009)  547 [arXiv:0901.3485 [hep-ph]].

%
\bibitem{Das:2010ww}
  D.~Das, U.~Ellwanger,
  JHEP {\bf 1009 } (2010)  085  [arXiv:1007.1151 [hep-ph]].
%
\bibitem{Gunion:2010dy}
  J.~F.~Gunion, A.~V.~Belikov, D.~Hooper,
  arXiv:1009.2555 [hep-ph].
%
\bibitem{Draper:2010ew}
  P.~Draper, T.~Liu, C.~E.~M.~Wagner, L.~T.~M.~Wang and H.~Zhang,
  Phys.\ Rev.\ Lett.\  {\bf 106} (2011) 121805
  [arXiv:1009.3963 [hep-ph]].
%
\bibitem{Kappl:2010qx}
  R.~Kappl, M.~Ratz, M.~W.~Winkler,
  Phys.\ Lett.\  {\bf B695 } (2011)  169 [arXiv:1010.0553 [hep-ph]].
%
\bibitem{Bottino:2002ry}
  A.~Bottino, N.~Fornengo, S.~Scopel,
  Phys.\ Rev.\  {\bf D67 } (2003)  063519
  [hep-ph/0212379].

\bibitem{Bottino:2004qi}
  A.~Bottino, F.~Donato, N.~Fornengo, S.~Scopel,
  Phys.\ Rev.\  {\bf D70 } (2004)  015005
 [hep-ph/0401186].

\bibitem{Bottino:2008mf}
  A.~Bottino, F.~Donato, N.~Fornengo, S.~Scopel,
  Phys.\ Rev.\  {\bf D78 } (2008)  083520 [arXiv:0806.4099 [hep-ph]].

\bibitem{Buras:2002vd}
  A.~J.~Buras, P.~H.~Chankowski, J.~Rosiek, {\L}.~S{\l}awianowska,
  Nucl.\ Phys.\  {\bf B659} (2003)  3
  [hep-ph/0210145].

\bibitem{Isidori:2006pk}
  G.~Isidori, P.~Paradisi,
  Phys.\ Lett.\  {\bf B639 } (2006)  499-507
  [hep-ph/0605012].

\bibitem{Barenboim:2007sk}
  G.~Barenboim, P.~Paradisi, O.~Vives, E.~Lunghi, W.~Porod,
  JHEP {\bf 0804 } (2008)  079
  [arXiv:0712.3559 [hep-ph]].

\bibitem{Eriksson:2008cx}
  D.~Eriksson, F.~Mahmoudi, O.~St{\r a}l,
  JHEP {\bf 0811 } (2008)  035 [arXiv:0808.3551 [hep-ph]].


\bibitem{Altmannshofer:2009ne}
  W.~Altmannshofer, A.~J.~Buras, S.~Gori, P.~Paradisi, D.~M.~Straub,
  Nucl.\ Phys.\  {\bf B830 } (2010)  17 [arXiv:0909.1333 [hep-ph]].

\bibitem{Altmannshofer:2010zt}
  W.~Altmannshofer, D.~M.~Straub,
  JHEP {\bf 1009 } (2010)  078 [arXiv:1004.1993 [hep-ph]].


\bibitem{Hou:1992sy}
  W.~-S.~Hou,
  Phys.\ Rev.\  {\bf D48 } (1993)  2342-2344.

\bibitem{Akeroyd:2003zr}
  A.~G.~Akeroyd and S.~Recksiegel,
  J.\ Phys.\ G {\bf 29} (2003) 2311
  [arXiv:hep-ph/0306037].

%
\bibitem{Antonelli:2008jg}
  M.~Antonelli {\it et al.} [FlaviaNet Working Group on Kaon Decays Collaboration],
arXiv:0801.1817 [hep-ph].
%
\bibitem{Nakamura:2010zzi}
  K. Nakamura {\it et al.} [Particle Data Group Collaboration],
  J.\ Phys.\ G {\bf G37 } (2010)  075021.
%
\bibitem{Kamenik:2008tj}
  J.~F.~Kamenik, F.~Mescia,
  Phys.\ Rev.\  {\bf D78 } (2008)  014003  [arXiv:0802.3790 [hep-ph]].
%
\bibitem{Akeroyd:2009tn}
  A.~G.~Akeroyd, F.~Mahmoudi,
  JHEP {\bf 0904 } (2009)  121 [arXiv:0902.2393 [hep-ph]].


\bibitem{Antonelli:2010yf}
  M.~Antonelli, V.~Cirigliano, G.~Isidori, F.~Mescia, M.~Moulson, H.~Neufeld, 
E.~Passemar, M.~Palutan {\it et al.},
  Eur.\ Phys.\ J.\  {\bf C69 } (2010)  399 [arXiv:1005.2323 [hep-ph]].

%
\bibitem{Aubert:2007dsa}
  B.~Aubert {\it et al.} [BABAR Collaboration],
  Phys.\ Rev.\ Lett.\  {\bf 100 } (2008)  021801 [arXiv:0709.1698 [hep-ex]].
%
\bibitem{Onyisi:2009th}
  P.~U.~E.~Onyisi {\it et al.} [CLEO Collaboration],
  Phys.\ Rev.\  {\bf D79 } (2009)  052002 [arXiv:0901.1147 [hep-ex]].
%
\bibitem{Alexander:2009ux}
  J.~P.~Alexander {\it et al.} [CLEO Collaboration],
  Phys.\ Rev.\  {\bf D79 } (2009)  052001 [arXiv:0901.1216 [hep-ex]].
%
\bibitem{Mahmoudi:2010xp}
  F.~Mahmoudi, J.~Rathsman, O.~St{\r a}l, L.~Zeune,
  arXiv:1012.4490 [hep-ph].

\bibitem{Bertolini:1990if}
  S.~Bertolini, F.~Borzumati, A.~Masiero, G.~Ridolfi,
  Nucl.\ Phys.\  {\bf B353 } (1991)  591-649.

\bibitem{Ciuchini:1997xe}
  M.~Ciuchini, G.~Degrassi, P.~Gambino, G.~F.~Giudice,
  Nucl.\ Phys.\  {\bf B527 } (1998)  21-43
  [hep-ph/9710335].


\bibitem{Barbieri:1993av}
  R.~Barbieri and G.~F.~Giudice,
  Phys.\ Lett.\  B {\bf 309} (1993) 86
  [arXiv:hep-ph/9303270].

\bibitem{Okada:1993sx}
  Y.~Okada,
  Phys.\ Lett.\  B {\bf 315} (1993) 119
  [arXiv:hep-ph/9307249].

\bibitem{Garisto:1993jc}
  R.~Garisto and J.~N.~Ng,
  Phys.\ Lett.\  B {\bf 315} (1993) 372
  [arXiv:hep-ph/9307301].

\bibitem{Ciuchini:1998xy}
  M.~Ciuchini, G.~Degrassi, P.~Gambino, G.~F.~Giudice,
  Nucl.\ Phys.\  {\bf B534 } (1998)  3-20
  [hep-ph/9806308].

\bibitem{Carena:2000uj}
  M.~S.~Carena, D.~Garcia, U.~Nierste, C.~E.~M.~Wagner,
  Phys.\ Lett.\  {\bf B499 } (2001)  141-146
  [hep-ph/0010003].

\bibitem{Degrassi:2000qf}
  G.~Degrassi, P.~Gambino, G.~F.~Giudice,
  JHEP {\bf 0012 } (2000)  009
  [hep-ph/0009337].

\bibitem{Belanger:2010cd}
  G.~B{\'e}langer, M.~Kakizaki, E.~K.~Park, S.~Kraml, A.~Pukhov,
  JCAP {\bf 1011 } (2010)  017 [arXiv:1008.0580 [hep-ph]].


\bibitem{ALEPH:2005ema}
  S.~Schael {\it et al.} [ALEPH and DELPHI and L3 and OPAL and SLD and LEP Electroweak Working Group and SLD Electroweak Group and SLD Heavy Flavour Group Collaborations],
  Phys.\ Rept.\  {\bf 427 } (2006)  257 [hep-ex/0509008].
\bibitem{Abbiendi:2003sc}
  G.~Abbiendi {\it et al.} [OPAL Collaboration],
  Eur.\ Phys.\ J.\  {\bf C35 } (2004)  1 [hep-ex/0401026].

\bibitem{Barberio:2008fa}
  E.~Barberio {\it et al.} [Heavy Flavor Averaging Group Collaboration],
  arXiv:0808.1297 [hep-ex].

\bibitem{Abazov:2010fs}
  V.~M.~Abazov {\it et al.} [D0 Collaboration],
  Phys.\ Lett.\  {\bf B693 } (2010)  539 [arXiv:1006.3469 [hep-ex]].


\bibitem{Djouadi:2002ze}
  A.~Djouadi, J.~L.~Kneur and G.~Moultaka,
  Comput.\ Phys.\ Commun.\  {\bf 176} (2007) 426
  [arXiv:hep-ph/0211331].


\bibitem{micro}
  G.~B{\'e}langer, F.~Boudjema, A.~Pukhov, A.~Semenov,
  Comput.\ Phys.\ Commun.\  {\bf 176 } (2007)  367 [hep-ph/0607059];
  G.~B{\'e}langer, F.~Boudjema, P.~Brun, A.~Pukhov, S.~Rosier-Lees, P.~Salati, A.~Semenov,
  Comput.\ Phys.\ Commun.\  {\bf 182} (2011)  842 [arXiv:1004.1092 [hep-ph]].
%
\bibitem{Mahmoudi:2008tp}
  F.~Mahmoudi,
  Comput.\ Phys.\ Commun.\  {\bf 180} (2009)  1579 [arXiv:0808.3144 [hep-ph]].

\bibitem{Ellis:1983er}
  J.~R.~Ellis, J.~M.~Fr{\`e}re, J.~S.~Hagelin, G.~L.~Kane, S.~T.~Petcov,
  Phys.\ Lett.\  {\bf B132 } (1983)  436.

\bibitem{Bartl:1986hp}
  A.~Bartl, H.~Fraas, W.~Majerotto,
  Nucl.\ Phys.\  {\bf B278 } (1986)  1.

\bibitem{Barbieri:1987hb}
  R.~Barbieri, G.~Gamberini, G.~F.~Giudice, G.~Ridolfi,
  Phys.\ Lett.\  {\bf B195 } (1987)  500.

\bibitem{Heinemeyer:2007bw}
  S.~Heinemeyer, W.~Hollik, A.~M.~Weber, G.~Weiglein,
  JHEP {\bf 0804 } (2008)  039
  [arXiv:0710.2972 [hep-ph]].

\bibitem{Djouadi:2005gj}
  A.~Djouadi,
  Phys.\ Rept.\  {\bf 459 } (2008)  1  [hep-ph/0503173].

\bibitem{Schael:2006cr} 
  S.~Schael {\it et al.} [ALEPH and DELPHI and
  L3 and OPAL and LEP Working Group for Higgs Boson Searches
  Collaborations],
  Eur.\ Phys.\ J.\  {\bf C47 } (2006)  547-587 [hep-ex/0602042].


\bibitem{cms}
CMS Collaboration, ``The CMS physics reach for searches at 7 TeV'', public note, 
CMS-NOTE-2010-008, CERN-CMS-NOTE-2010-008.

\bibitem{tevatron}
D.~Benjamin {\it et al.} [Tevatron New Phenomena \& Higgs Working Group Collaboration],
  arXiv:1003.3363 [hep-ex].


\bibitem{Khachatryan:2011tk}
  V.~Khachatryan {\it et al.} [CMS Collaboration],
  Phys.\ Lett.\  {\bf B698 } (2011)  196-218
  [arXiv:1101.1628 [hep-ex]].

\bibitem{daCosta:2011qk}
  G.~Aad {\it et al.} [ATLAS Collaboration],
  arXiv:1102.5290 [hep-ex].

\bibitem{Scopel:2011qt}
  S.~Scopel, S.~Choi, N.~Fornengo, A.~Bottino,
  arXiv:1102.4033 [hep-ph].


\bibitem{cdf}
  T.~Aaltonen {\it et al.} [CDF Collaboration],
  CDF public note 9892.

\bibitem{Aaij:2011rj}
  R. Aaij {\it et al.} [LHCb Collaboration],
  arXiv:1103.2465 [hep-ex].

\bibitem{Jegerlehner:2011ti}
  F.~Jegerlehner and R.~Szafron,
  Eur.\ Phys.\ J.\  C {\bf 71} (2011) 1632
  [arXiv:1101.2872 [hep-ph]].

\bibitem{Ahmed:2009zw}
  Z.~Ahmed {\it et al.} [CDMS-II Collaboration],
  Science {\bf 327 } (2010)  1619-1621 [arXiv:0912.3592 [astro-ph.CO]].

\bibitem{Angle:2009xb}
  J.~Angle {\it et al.} [XENON10 Collaboration],
  Phys.\ Rev.\  {\bf D80 } (2009)  115005 [arXiv:0910.3698 [astro-ph.CO]].


\bibitem{Ahmed:2010wy}
  Z.~Ahmed {\it et al.} [CDMS-II Collaboration],
  Phys.\ Rev.\ Lett.\  {\bf 106 } (2011)  131302  [arXiv:1011.2482 [astro-ph.CO]].


\bibitem{Aprile:2010um}
  E.~Aprile {\it et al.} [XENON100 Collaboration],
  Phys.\ Rev.\ Lett.\  {\bf 105 } (2010)  131302 [arXiv:1005.0380 [astro-ph.CO]].

\bibitem{Aprile:2011hx}
  E.~Aprile {\it et al.} [XENON100 Collaboration],
  arXiv:1103.0303 [hep-ex].

\bibitem{Chatrchyan:2011nx}
  S.~Chatrchyan {\it et al.}  [CMS Collaboration],
  arXiv:1104.1619 [hep-ex].

\bibitem{baglio}
  J.~Baglio,
  arXiv:1105.1085 [hep-ph];
A.~Djouadi, ``Implications of the first Supersymmetric Higgs
searches at the LHC'', Talk given at Planck 2011 conference, 30. May - 3. June 2011, Lisbon, Portugal.

  

\end{thebibliography}
\end{document}